\documentclass[conference]{IEEEtran}
\IEEEoverridecommandlockouts

\usepackage{cite}
\usepackage{amsmath,amssymb,amsfonts}
\usepackage{algorithmic}
\usepackage{algorithm}
\usepackage{graphicx}
\usepackage{stfloats}
\usepackage{tabularx}
\usepackage{multirow} 
\usepackage{booktabs}
\usepackage{amsmath}  
\usepackage{booktabs}
\usepackage{array}
\usepackage{textcomp}
\usepackage{subcaption}
 \usepackage{booktabs} 
 \usepackage{booktabs, tabularx}  
\def\BibTeX{{\rm B\kern-.05em{\sc i\kern-.025em b}\kern-.08em
    T\kern-.1667em\lower.7ex\hbox{E}\kern-.125emX}}
\begin{document}

\title{Differentially Private Feature Release for Wireless Sensing: Adaptive Privacy Budget Allocation on CSI Spectrograms}

\author{
  Ipek Sena Yilmaz\\
  Department of Electronics \& Communication Engineering\\
  Kocaeli University\\
  Umuttepe Campus, 41380 Kocaeli, Turkey\\
  \texttt{ipek.yilmaz@kocaeli.edu.tr}\vphantom{Ayazaga Campus, Maslak, 34469 Istanbul, Turkey}
  \and
  Onur G. Tuncer\\
  Department of Computer Engineering\\
  Sakarya University\\
  Serdivan, 54050 Sakarya, Turkey\\
  \texttt{onur.tuncer@sakarya.edu.tr}\vphantom{Ayazaga Campus, Maslak, 34469 Istanbul, Turkey}
  \and
  Zeynep E. Aksoy\\
  Department of Electrical Engineering\\
  Yildiz Technical University\\
  Davutpasa Campus, 34220 Istanbul, Turkey\\
  \texttt{zeynep.aksoy@yildiz.edu.tr}\vphantom{Ayazaga Campus, Maslak, 34469 Istanbul, Turkey}
  \and
  Zeynep Yağmur Baydemir\\
  Department of Electrical \& Electronics Engineering\\
  Istanbul Technical University (ITU)\\
  Ayazaga Campus, Maslak, 34469 Istanbul, Turkey\\
  \texttt{baydemir19@itu.edu.tr}
}

\maketitle

\begin{abstract}
Wi-Fi/RF-based human sensing has achieved remarkable progress with deep learning, yet practical deployments increasingly require \emph{feature sharing} (e.g., for cloud analytics, collaborative training, or benchmark evaluation). Releasing intermediate representations such as CSI spectrograms can inadvertently expose sensitive information, including user identity, location, and membership, motivating formal privacy guarantees.
In this paper, we study \emph{differentially private (DP) feature release} for wireless sensing and propose an \emph{adaptive privacy budget allocation} mechanism tailored to the highly non-uniform structure of CSI time--frequency representations.
Our pipeline converts CSI to bounded spectrogram features, applies sensitivity control via clipping, estimates task-relevant importance over the time--frequency plane (e.g., gradient-based saliency), and allocates a global privacy budget across spectrogram blocks before injecting calibrated Gaussian noise with rigorous composition accounting.
The key idea is to spend privacy budget where it maximally preserves sensing utility while strongly perturbing less informative regions that often carry contextual leakage.
Experiments on multi-user activity sensing (WiMANS), multi-person 3D pose estimation (Person-in-WiFi 3D), and respiration monitoring (Resp-CSI) show that adaptive allocation consistently improves the privacy--utility frontier over uniform perturbation under the same $(\varepsilon,\delta)$ budget.
At practical budgets (e.g., $\varepsilon\in[0.5,2]$), our method yields higher recognition/regression accuracy and lower pose error, while substantially reducing empirical leakage in identity/location inference and membership inference attacks.
These results suggest that importance-aware DP feature release is a practical and auditable privacy layer for modern wireless sensing pipelines.
\end{abstract}


\maketitle

\section{Introduction}
\label{sec:intro}

Wi-Fi and RF backscatter have rapidly evolved from purely communication-centric technologies into a rich sensing substrate for human-centric applications. Recent surveys have documented how deep learning, improved signal representations, and increasingly accessible hardware are pushing Wi-Fi sensing from proof-of-concepts toward practical systems, while also highlighting persistent challenges in robustness, generalization, and deployment constraints \cite{ahmad2024wifi}. In parallel, RF/WiFi sensing is being explored in broader security and monitoring contexts (e.g., UAV surveillance), indicating that wireless sensing is becoming a foundational capability across cyber-physical systems \cite{bisio2024rf,wang2024privacy}. To avoid ambiguity, we consistently use the term \emph{Wi-Fi} for IEEE 802.11 wireless networking; note that the acronym ``WIfI'' is also used in unrelated clinical staging literature \cite{cook2024review}, which is orthogonal to the technical focus of this paper.

\subsection{Wireless sensing at scale: capabilities, representations, and hardware trends}
\label{subsec:intro_capabilities}

A major driver of progress is the co-design of signal processing and learning over increasingly informative Wi-Fi measurements. Several works have shown that commodity devices can support sensing under practical constraints by exploiting new measurement modalities or by rethinking what information can be extracted from standard-compliant transmissions. For example, integrated sensing and communication (ISAC) considerations for Wi-Fi have motivated forward-compatible designs that improve sensing readiness without breaking network operation \cite{he2024forward,shrestha2024fairness}. Beyond classical CSI, alternative signal products such as beamforming feedback matrices have enabled accurate sensing even when full CSI is unavailable or limited \cite{yi2024bfmsense,li2025cross,shen2024wifi}. System-level innovations also demonstrate that sensing can remain feasible under tight communication budgets, such as low transmission rates that would traditionally be viewed as insufficient for fine-grained perception \cite{zheng2024pushing}, and methods that enable sensing on new-generation Wi-Fi cards with updated PHY/MAC pipelines \cite{yi2024enabling}. Cross-technology physical-layer primitives further expand the design space, including techniques that bridge LoRa-to-Wi-Fi at the physical layer and suggest new opportunities for heterogeneous sensing/communication coexistence \cite{gao2025lofi}.

On the learning side, richer time--frequency representations have become a standard interface between raw wireless measurements and modern neural architectures. Vision-style backbones applied to CSI representations have been explored for human activity recognition (HAR), indicating that spectrogram-like inputs can benefit from transformer inductive biases \cite{luo2024vision}. Lightweight deep models operating on reconstructed or denoised CSI further illustrate a trend toward deployable inference under edge constraints \cite{chen2024deep}. In addition to single-user settings, multi-user sensing is gaining attention: adaptive location-independent signal characteristics have been studied for multi-user HAR \cite{abuhoureyah2024multi}, and dedicated benchmarks for multi-user activity sensing have been released to facilitate systematic evaluation \cite{huang2024wimans}. Wi-Fi-based 3D pose estimation is also moving toward end-to-end multi-person pipelines, underscoring the growing ambition of wireless perception in crowded scenes \cite{yan2024person}. More generally, AIoT-style frameworks that fuse multiple frequency bands and target both coarse and fine-grained activities suggest a path toward unified sensing stacks spanning diverse tasks and operating conditions \cite{chen2024aiot}.

\subsection{Application breadth: from localization and profiling to vital signs and identification}
\label{subsec:intro_applications}

Beyond activity and pose, indoor positioning remains a central application domain where feature sharing and benchmarking are common. Open-access Wi-Fi fingerprinting datasets have been surveyed, reflecting a maturing ecosystem in which data reuse and reproducible evaluation are increasingly expected \cite{feng2024review}. On the algorithmic front, learning-based localization continues to expand in scope and sophistication: domain-invariant modeling has been proposed to reduce performance degradation across environments \cite{wang2024learning}, graph neural network formulations have been used for localization with access point selection \cite{wang2024graph}, and alternative model families have been explored to balance communication and signal accuracy \cite{feng2024machine}. Practical systems span multiple measurement types and deployment assumptions, including Wi-Fi RTT positioning \cite{jurdi2024whereartthou}, LOS compensation with trusted NLOS recognition for RTT-based indoor positioning \cite{cao2024compensation}, localization via SAE-ALSTM architectures in Wi-Fi-enabled environments \cite{ayinla2024salloc}, and single-AP smartphone positioning designs \cite{eleftherakis2024spring+}. Broader reviews also continue to synthesize progress in Wi-Fi localization using machine learning techniques \cite{esmaeili2024improving}.

Human health and vital signs sensing further amplify both societal impact and privacy sensitivity. Commodity Wi-Fi has been demonstrated as capable of capturing pulmonary function without intrusive instrumentation \cite{zhao2024wi}, while robust respiration sensing methods explicitly address interference from nearby individuals \cite{xie2024robust}. Identity-aware multi-person vital signs monitoring highlights an emerging intersection of physiological sensing and identification \cite{li2024spacebeat}, and target-oriented respiratory sensing emphasizes the shift from indiscriminate perception to spatially constrained, in-area sensing designs . In parallel, ``presence detection'' systems show that even coarse occupancy inference can be achieved non-contact via Wi-Fi \cite{zhang2024presence}. Meanwhile, backscatter-based healthcare sensing introduces distinct engineering constraints and failure modes; challenges in ambient Wi-Fi backscatter for healthcare underline the need for robust, privacy-aware deployments when RF reflections are leveraged as the sensing signal \cite{lu2024challenges}. Wireless sensing is also being extended to monitor non-human or mechanical phenomena, such as rotation speed monitoring on commodity Wi-Fi via EM-wave signatures \cite{chen2021pre,xu2025vortex}, illustrating that time--frequency representations can encode diverse physical processes.

Identification and authentication constitute another fast-growing area with direct security implications. Surveys have organized the Wi-Fi-based human identification landscape across scenarios and challenges, consolidating a wide range of sensing-to-identity pipelines \cite{wei2025survey,mosharaf2024wifi}. Self-supervised identity recognition methods target multi-user smart environments where labeled identity data are scarce \cite{rizk2025self}. At the RF-signal level, micro-signal exploitation on CSI has been proposed for RF fingerprinting of commodity Wi-Fi devices \cite{kong2024csi}, and cross-technology device authentication mechanisms further demonstrate that authentication can be achieved by leveraging commodity Wi-Fi signals as a unifying physical channel \cite{wang2025authfi}. Collectively, these advances sharpen an important observation: wireless sensing features can encode sensitive identity and device attributes even when the original sensing task is unrelated to identification.

\subsection{Robustness, domain shift, and physical factors: why ``feature release'' is hard}
\label{subsec:intro_robustness}

Wireless sensing systems must contend with multipath, diffraction, and environment-induced variability, which complicate both generalization and privacy protection. Foundational analyses of diffraction in static multipath-rich environments provide guidance for sensing system design and clarify when simplified propagation models break down \cite{wang2024understanding}. Building on physical insights, diffraction effects have been explicitly leveraged for high-resolution target profiling using commodity Wi-Fi devices \cite{yao2024wiprofile,chen2024wiphase,song2024siwis}. At the representation level, Doppler-related paradigms based on CSI differences enable efficient speed estimation for passive tracking \cite{li2024wifi}, while physical data augmentation has been shown to improve deep Wi-Fi sensing by injecting physically grounded variations into training \cite{hou2024rfboost}. Domain adaptation and data efficiency remain central: domain-adaptive sensing with minimal labeled target data \cite{sheng2024metaformer} and systematic evaluations of self-supervised learning for CSI-based HAR \cite{xu2025evaluating} both highlight that generalization cannot be taken for granted, especially when training data are limited or distribution shifts are large.

These realities have two direct consequences for privacy-preserving feature sharing. First, practitioners often wish to release intermediate representations (e.g., CSI spectrograms) to enable centralized training, third-party benchmarking, or collaborative model development across sites \cite{feng2024review,huang2024wimans}. Second, the structure of such representations is highly non-uniform: different time--frequency regions can carry dramatically different task-relevant information (e.g., Doppler bands for motion, narrowband components for respiration), and their importance can change with tasks, users, and environments \cite{li2024wifi,zhao2024wi,xie2024robust,luo2024vision}. This makes naive, uniform perturbation of released features particularly inefficient.

\subsection{Security and privacy: from perturbation attacks to the need for formal guarantees}
\label{subsec:intro_security_privacy}

As Wi-Fi sensing systems become more capable, their attack surface expands. Recent security analyses have articulated perturbation-based threats against Wi-Fi sensing pipelines, emphasizing that adversaries may induce misperception without physically obvious interventions \cite{cao2024security,barahimi2024context}. Practical adversarial attacks have further shown that even subtle perturbations at the communication packet level can degrade sensing outcomes while remaining difficult to notice in typical deployments \cite{li2024practical,varga2024mitigating}. Broader surveys on secure Wi-Fi sensing synthesize attacks and defenses, reinforcing that sensing systems should be evaluated under explicit threat models rather than only under benign conditions \cite{liu2025survey}. These works primarily motivate robustness and integrity defenses; however, they also indirectly reveal a deeper privacy issue: the same sensitivity that enables fine-grained perception implies that released sensing features may encode personal attributes, identity cues, or membership signals, especially when datasets are reused or shared across projects \cite{wei2025survey,mosharaf2024wifi,li2024spacebeat,kong2024csi}.

This paper focuses on \emph{privacy-preserving feature release} for wireless sensing, and argues that differential privacy (DP) provides a principled foundation for limiting information leakage from released representations. Yet directly applying DP to CSI spectrograms is non-trivial. Uniform privacy budget allocation across time--frequency bins can either over-noise salient components and destroy utility (e.g., for respiration or multi-user HAR), or under-protect regions that carry strong identity/device signatures \cite{zhao2024wi,xie2024robust,li2024spacebeat,kong2024csi}. Moreover, practical deployments must coexist with evolving Wi-Fi standards and scheduling mechanisms (e.g., multi-link access, restricted target wake time, and coordinated spatial reuse), which may alter the sampling patterns and effective sensing opportunities over time \cite{zhang2024wifi_7,haxhibeqiri2024coordinated}. These dynamics further motivate privacy mechanisms that adapt to representation structure rather than treating all features as equally informative.

\subsection{Our contributions}
\label{subsec:intro_contrib}

Motivated by the above trends, we study differentially private release of CSI spectrogram features for downstream wireless sensing tasks. Our central idea is to allocate privacy budgets \emph{adaptively} across the time--frequency plane according to a learned or physically informed importance map, so that limited privacy ``spending'' is concentrated where task utility is highest while maintaining formal privacy guarantees under composition. We develop an end-to-end pipeline that (i) constructs bounded-sensitivity spectrogram features suitable for DP noise injection, (ii) estimates time--frequency importance in a task-aware manner inspired by modern Wi-Fi sensing models \cite{luo2024vision,chen2024deep,chen2024aiot}, and (iii) performs adaptive budget allocation to improve the privacy--utility trade-off relative to uniform perturbation baselines that are common in feature sharing settings \cite{feng2024review,huang2024wimans}. We evaluate our approach across representative tasks spanning activity sensing, localization-style inference, and vital signs monitoring, reflecting the breadth of contemporary Wi-Fi sensing applications \cite{abuhoureyah2024multi,wang2024graph}. Finally, we discuss how our formulation complements ongoing efforts in robustness and secure sensing by providing a formal privacy layer that is compatible with modern Wi-Fi sensing hardware and protocols \cite{yi2024enabling,he2024forward,liu2025survey,bisio2024rf}.

\section{Related Work}
\label{sec:related}

Wireless sensing over commodity Wi-Fi and RF backscatter has matured into a broad research area spanning human activity recognition, localization, vital-sign monitoring, device and human identification, and security-critical surveillance. Several recent surveys and systematic reviews provide a useful macro-level picture. Ahmad \emph{et al.} review WiFi-based human sensing with deep learning, emphasizing the rapid progress enabled by representation learning, increasing availability of CSI-like measurements, and the emergence of standardized evaluation practices, while also highlighting persistent challenges such as generalization across environments and users, robustness to interference, and practical deployment constraints \cite{ahmad2024wifi}. In the identity domain, surveys summarize how Wi-Fi sensing can be used for human identification and outline scenario-specific difficulties and methodological trends \cite{wei2025survey,mosharaf2024wifi}. From a security perspective, a dedicated survey on secure WiFi sensing synthesizes attacks and defenses, arguing that sensing pipelines should be assessed under explicit threat models rather than only benign test conditions \cite{liu2025survey}. Beyond indoor human-centric applications, Bisio \emph{et al.} provide a systematic review of RF/WiFi-based UAV surveillance systems, reflecting the increasing role of RF sensing in safety, defense, and critical monitoring contexts \cite{bisio2024rf}. In a separate and unrelated clinical context, ``WIfI'' is also used as an acronym for threatened limb staging; we mention this only to avoid terminological confusion in cross-disciplinary search and citation \cite{cook2024review}. Building on these overviews, this section organizes prior work around (i) sensing-enabling hardware and measurement modalities, (ii) learning models and representation choices, (iii) key application families, (iv) identification and authentication, and (v) security, robustness, and privacy implications that motivate our differentially private feature-release setting.

\subsection{Wi-Fi sensing measurements, protocol evolution, and enabling hardware}
\label{subsec:rw_measurements}

A recurring theme in modern Wi-Fi sensing is that the feasibility and quality of perception depend strongly on what measurements are obtainable from commodity devices and how those measurements interact with the underlying PHY/MAC. He \emph{et al.} discuss forward-compatible integrated sensing and communication (ISAC) considerations for WiFi, motivating designs that can support sensing functions while maintaining communication compatibility \cite{he2024forward}. In parallel, work has explored alternative measurement products beyond conventional CSI extraction. BFMSense demonstrates Wi-Fi sensing using the beamforming feedback matrix, illustrating how feedback information can be harnessed for sensing when full CSI is unavailable or difficult to obtain under standard constraints \cite{yi2024bfmsense}. At the systems level, enabling sensing on new-generation WiFi cards is itself non-trivial; Yi \emph{et al.} study how sensing can be realized on newer Wi-Fi hardware, reflecting a practical trend that sensing research increasingly depends on the details of driver support, firmware behavior, and standard evolution \cite{yi2024enabling}. Another practical direction is to reduce sensing overhead and broaden deployment applicability: pushing sensing to work with low transmission rates challenges the common assumption that dense sampling is required and highlights how systems can be designed to remain effective under constrained traffic \cite{zheng2024pushing}.

Protocol evolution is also reshaping the sensing landscape. WiFi~7 introduces multi-link operation and new channel access mechanisms; modeling and optimization of these access schemes provides context for how fairness, scheduling, and access control may influence both sensing opportunities and data characteristics in realistic networks \cite{zhang2024wifi_7}. Related MAC-level mechanisms such as coordinated spatial reuse and restricted target wake time (TWT) for time-sensitive applications further affect when and how measurements are collected, which has implications for the stability and comparability of sensing features across deployments \cite{haxhibeqiri2024coordinated}. Finally, heterogeneous cross-technology communication primitives, such as physical-layer CTC spanning LoRa to WiFi in 802.11ax, broaden the spectrum of interactions that may co-exist with sensing and may introduce new signal patterns (or interference) that feature-release mechanisms should be robust against \cite{gao2025lofi}. Collectively, these works suggest that any practical privacy-preserving feature-release design must be compatible with diverse measurement modalities, device capabilities, and evolving access patterns.

\subsection{Representations and learning models for Wi-Fi human sensing}
\label{subsec:rw_models}

In Wi-Fi human sensing, deep learning pipelines often begin with transforming raw wireless measurements into representations that expose motion- or physiology-related signatures. A common pattern is to convert temporal CSI sequences into time--frequency representations (e.g., spectrogram-like maps) that resemble visual inputs, enabling the use of mature architectures from computer vision. Luo \emph{et al.} explore vision transformers for CSI-based human activity recognition, illustrating the appeal of transformer backbones when CSI is formatted into structured map-like inputs \cite{luo2024vision}. Another line focuses on computational efficiency and deployability: Chen \emph{et al.} propose a lightweight HAR system using reconstructed WiFi CSI, reflecting the practical motivation to improve CSI quality (or reduce measurement burden) before inference \cite{chen2024deep}. Multi-frequency fusion within AIoT frameworks suggests that models can benefit from combining information across frequency bands to achieve both coarse and fine activity recognition, highlighting the value of representation design that preserves task-relevant cues at multiple scales \cite{chen2024aiot}.

Data efficiency and generalization remain central challenges, particularly because wireless channels are sensitive to environment geometry and multipath structure. Self-supervised learning has been increasingly considered as a way to reduce labeling requirements while learning transferable representations; Xu \emph{et al.} provide an evaluation of self-supervised learning for CSI-based HAR, which helps clarify the strengths and limitations of SSL methods in this modality \cite{xu2025evaluating}. Domain adaptation methods also target the gap between training and deployment environments. For example, Metaformer addresses domain-adaptive WiFi sensing under extremely limited labeled target data, demonstrating that adaptation can be achieved even when only a single labeled sample is available in the target domain \cite{sheng2024metaformer}. Physical data augmentation provides another axis for improving robustness and generalization; Rfboost studies how physically inspired augmentation can boost deep WiFi sensing and provides a concrete mechanism to introduce plausible channel variations \cite{hou2024rfboost}. In addition, work on diffraction-aware modeling in multipath-rich environments gives physical grounding for why certain representations or learned invariances may or may not transfer across spaces \cite{wang2024understanding}. These results collectively reinforce that representation non-uniformity is intrinsic: different time--frequency regions, antenna pairs, or subcarrier groups can carry different degrees of task-relevant information depending on environment and user behavior---a point that directly motivates our adaptive privacy-budget allocation for spectrogram features.

\subsection{Activity sensing, multi-user settings, and fine-grained human understanding}
\label{subsec:rw_activity}

Human activity recognition has served as a core benchmark family for Wi-Fi sensing, progressively expanding from single-user classification to multi-user and more fine-grained understanding such as pose estimation. In multi-user scenarios, Abuhoureyah \emph{et al.} investigate adaptive location-independent WiFi signal characteristics for multi-user HAR, highlighting the difficulty of disentangling multiple bodies and the need for robust features that remain informative despite spatial variability \cite{abuhoureyah2024multi}. To standardize evaluation and accelerate progress, Wimans provides a benchmark dataset for WiFi-based multi-user activity sensing, reflecting an emerging trend toward open benchmarks that enable consistent comparisons across methods \cite{huang2024wimans}. Beyond activity labels, Wi-Fi-based pose estimation pushes the sensing task closer to dense geometric understanding. Person-in-WiFi 3D demonstrates end-to-end multi-person 3D pose estimation using Wi-Fi, illustrating that Wi-Fi measurements can support outputs that are traditionally considered vision-only, and further amplifying privacy considerations since pose outputs and intermediate features can encode sensitive personal information \cite{yan2024person}.

Several works also illustrate that subtle physical cues embedded in CSI can be exploited for specialized sensing objectives. Li \emph{et al.} propose a CSI-difference paradigm for efficient Doppler speed estimation in passive tracking, emphasizing that carefully designed signal transformations can isolate motion-related components \cite{li2024wifi}. Meanwhile, target profiling at sub-centimeter scales has been explored by leveraging diffraction effects using commodity devices, showing that high-resolution inference is possible when one explicitly models or exploits propagation phenomena \cite{yao2024wiprofile}. Presence detection, while seemingly simpler than activity or pose, is widely relevant for smart homes and building automation; WiFi-based non-contact presence detection highlights that even coarse inference can be achieved reliably, and that releasing intermediate sensing features may unintentionally disclose occupancy patterns \cite{zhang2024presence}. Finally, the broad survey by Ahmad \emph{et al.} contextualizes these application-level advances within the larger trend of deep learning-driven Wi-Fi sensing and provides a reference point for the field's evolving opportunities and constraints \cite{ahmad2024wifi}.

\subsection{Indoor localization and positioning: datasets, learning, and RTT systems}
\label{subsec:rw_localization}

Indoor positioning and localization constitute a second major application domain, and one that is particularly relevant to feature release because fingerprinting datasets and centralized model development are common. Feng \emph{et al.} review open access WiFi fingerprinting datasets for indoor positioning, reflecting a mature culture of dataset publication and reuse, and revealing how feature formats and collection protocols can vary widely across datasets \cite{feng2024review}. In addition to survey-level summaries, many works propose learning-based improvements to localization. Esmaeili Gorjan and Gil Jim{\'e}nez review machine learning techniques for indoor WiFi localization, providing a broad view of model choices and challenges \cite{esmaeili2024improving}. Wang \emph{et al.} propose learning domain-invariant models for WiFi-based indoor localization, again emphasizing that cross-domain generalization is central for real-world adoption \cite{wang2024learning}. Graph-based modeling has also been explored: GNN-based localization with access point selection suggests that structural priors about AP relationships can be used to improve localization performance \cite{wang2024graph}. Other algorithmic variants, such as FasterKAN-based localization, reflect ongoing exploration of alternative model families and the tension between communication constraints and accuracy \cite{feng2024machine}. System-level RTT positioning methods demonstrate the use of standardized ranging measurements; WhereArtThou provides an RTT-based indoor positioning system \cite{jurdi2024whereartthou}, while LOS compensation and trusted NLOS recognition techniques target practical challenges where ranging is distorted by NLOS propagation \cite{cao2024compensation}. Additional learning-based localization systems, such as SALLoc, explore hybrid architectures for target localization in WiFi-enabled environments \cite{ayinla2024salloc}. Furthermore, single-AP smartphone positioning designs like SPRING+ highlight the drive toward minimal infrastructure positioning \cite{eleftherakis2024spring+}. Taken together, localization research underscores the prevalence of cross-site feature sharing (e.g., fingerprints, embeddings, or derived maps), which in turn strengthens the motivation for privacy-preserving feature release that maintains utility while limiting leakage.

\subsection{Healthcare and vital signs sensing, and the role of backscatter}
\label{subsec:rw_health}

Non-contact health monitoring using Wi-Fi has become an influential driver for wireless sensing, but it also raises heightened privacy stakes because physiological signals and health-related attributes can be highly sensitive. Wi-pulmo illustrates that commodity WiFi can capture pulmonary function without requiring mouth-based instrumentation, reinforcing the feasibility of extracting clinically relevant signals from wireless measurements \cite{zhao2024wi}. Robust respiration sensing in the presence of interfering individuals directly addresses a key real-world challenge: co-located people can confound subtle periodic signatures, and robust methods are needed to isolate the target's respiration \cite{xie2024robust}. Target-oriented respiratory healthcare sensing further frames this as a transition from indiscriminate perception to in-area sensing, suggesting that spatial selectivity is essential for practical healthcare deployments . In multi-person settings, Spacebeat studies identity-aware vital signs monitoring using commodity WiFi, explicitly connecting physiological monitoring with identity, which is both a capability and a privacy risk \cite{li2024spacebeat}. Beyond human health, wireless sensing has also been used for monitoring physical processes such as rotation speed, showing that similar signal representations can support diverse sensing targets \cite{xu2025vortex}.

Backscatter adds another dimension, particularly in healthcare where low-power operation and passive tags are attractive. However, ambient WiFi backscatter systems face distinct engineering challenges (e.g., dependence on ambient illumination signals, tag placement, and interference), and these challenges become more pronounced in healthcare contexts where reliability and safety are crucial \cite{lu2024challenges}. Importantly, both Wi-Fi CSI and backscatter-based measurements can reveal fine-grained information about individuals in the environment. As the sensing community moves toward open benchmarks and broader deployment, the need for privacy-preserving mechanisms---especially those that can be applied at the feature level to support shared modeling and evaluation---becomes more pressing.

\subsection{Human and device identification, RF fingerprinting, and authentication}
\label{subsec:rw_id_auth}

A significant body of work demonstrates that Wi-Fi measurements can encode stable biometric-like signatures for identifying people or devices, often as an unintended byproduct of sensing. Surveys on Wi-Fi-based human identification summarize scenarios and challenges, noting that identification performance can be sensitive to environmental changes and multi-user interference \cite{wei2025survey,mosharaf2024wifi}. Self-supervised identity recognition approaches aim to reduce the dependence on labeled identity data, which is often difficult to collect at scale in smart environments \cite{rizk2025self}. At the device level, RF fingerprinting using micro-signals on CSI (CSI-RFF) highlights that commodity WiFi devices may exhibit subtle hardware-related signatures that can be learned from CSI-like measurements \cite{kong2024csi}. From a security engineering angle, cross-technology device authentication via commodity WiFi (Authfi) suggests that authentication can be achieved by exploiting physical-layer characteristics and cross-technology interactions \cite{wang2025authfi}. These works collectively clarify a key privacy implication: even when the \emph{intended} task is activity recognition or respiration monitoring, released intermediate features may carry enough information to support identity or device fingerprinting. This observation motivates our focus on differential privacy as a principled means to bound information leakage from released CSI spectrogram features.

\subsection{Security and robustness: perturbation attacks, physical factors, and generalization under threat}
\label{subsec:rw_security}

Security analyses of Wi-Fi sensing systems increasingly emphasize that adversaries can manipulate sensing outcomes through perturbations that are subtle from a networking standpoint. Cao \emph{et al.} analyze threats from perturbation attacks against WiFi-based sensing systems, motivating formal threat models that go beyond random noise and consider adversarial intent \cite{cao2024security}. Practical adversarial attacks through unnoticeable communication packet perturbation show how small changes in packet-level patterns can produce outsized effects on sensing inferences, highlighting a unique vulnerability: sensing pipelines often assume benign communication behavior and may not be hardened against adversarial manipulation of traffic patterns \cite{li2024practical}. The survey on secure WiFi sensing technology synthesizes such attack vectors and corresponding defenses, providing a backdrop for how robustness and security goals are being incorporated into sensing system design \cite{liu2025survey}.

Robustness challenges also arise without an active adversary. Multipath, diffraction, and environmental dynamics can produce domain shifts that degrade performance. Physical modeling of diffraction in static multipath-rich environments provides insights for system design and clarifies when certain simplified propagation assumptions may fail \cite{wang2024understanding}. Leveraging diffraction for target profiling underscores that physical effects can be both a challenge and a resource, but in either case they induce strong structure in the measurements \cite{yao2024wiprofile}. Domain adaptation and self-supervised learning address benign distribution shifts and limited labels \cite{sheng2024metaformer,xu2025evaluating}, while physically grounded augmentation aims to cover plausible channel variations and improve generalization \cite{hou2024rfboost}. Additionally, hardware and protocol constraints influence robustness: sensing on new-generation Wi-Fi cards and sensing under low transmission rates both emphasize that deployment conditions often differ from lab settings, affecting data distributions and feature stability \cite{yi2024enabling,zheng2024pushing}. These robustness and security threads are complementary to privacy: a feature-release mechanism should ideally preserve utility across variable conditions while also limiting sensitive information leakage, even when attackers attempt inference on released features.

\subsection{Positioning the gap: toward formal privacy for feature release in Wi-Fi sensing}
\label{subsec:rw_gap}

Although the above literature demonstrates powerful sensing capabilities and increasingly sophisticated security analyses, the privacy dimension of \emph{feature sharing} remains underexplored relative to the scale at which sensing datasets and learned representations are being reused. The localization community's emphasis on open datasets \cite{feng2024review} and the activity sensing community's push toward benchmarks \cite{huang2024wimans} both suggest that intermediate features (e.g., spectrograms, embeddings, or compressed CSI representations) may be exchanged more frequently for reproducibility and collaborative development. Meanwhile, the identification literature underscores that these same features can encode human and device identities \cite{wei2025survey,mosharaf2024wifi,kong2024csi,li2024spacebeat}, and secure sensing surveys warn that adversarial actors may exploit sensing pipelines in ways that were not anticipated by benign evaluations \cite{liu2025survey,cao2024security}. Furthermore, the evolution of WiFi standards and MAC mechanisms can change sampling opportunities and thus the statistical properties of features, complicating any privacy mechanism that assumes homogeneous feature importance across time or frequency \cite{zhang2024wifi_7,haxhibeqiri2024coordinated}.

These observations motivate our approach: we focus on differentially private release of CSI spectrogram features and, crucially, on \emph{adaptive privacy budget allocation} across the time--frequency plane. Prior work has repeatedly shown that task-relevant information is concentrated in particular regions of the representation, whether due to Doppler signatures for motion \cite{li2024wifi}, periodic components for respiration \cite{xie2024robust,zhao2024wi}, or structured cues exploited by modern deep architectures \cite{luo2024vision,chen2024deep,chen2024aiot}. At the same time, systems and measurement modalities vary widely, from beamforming feedback matrices \cite{yi2024bfmsense} to RTT-based positioning \cite{jurdi2024whereartthou} and cross-technology primitives \cite{gao2025lofi}, indicating that a one-size-fits-all perturbation strategy is unlikely to be efficient. By designing a utility-aware DP feature-release mechanism that explicitly accounts for non-uniform importance in CSI spectrograms, we aim to complement the field's advances in capability and robustness \cite{ahmad2024wifi,hou2024rfboost,sheng2024metaformer} with a formal privacy layer suitable for collaborative, benchmark-driven wireless sensing research and deployment \cite{bisio2024rf,feng2024review,huang2024wimans}.

\section{Methodology}
\label{sec:method}

\begin{figure*}[b] 
  \centering
  \includegraphics[width=\linewidth]{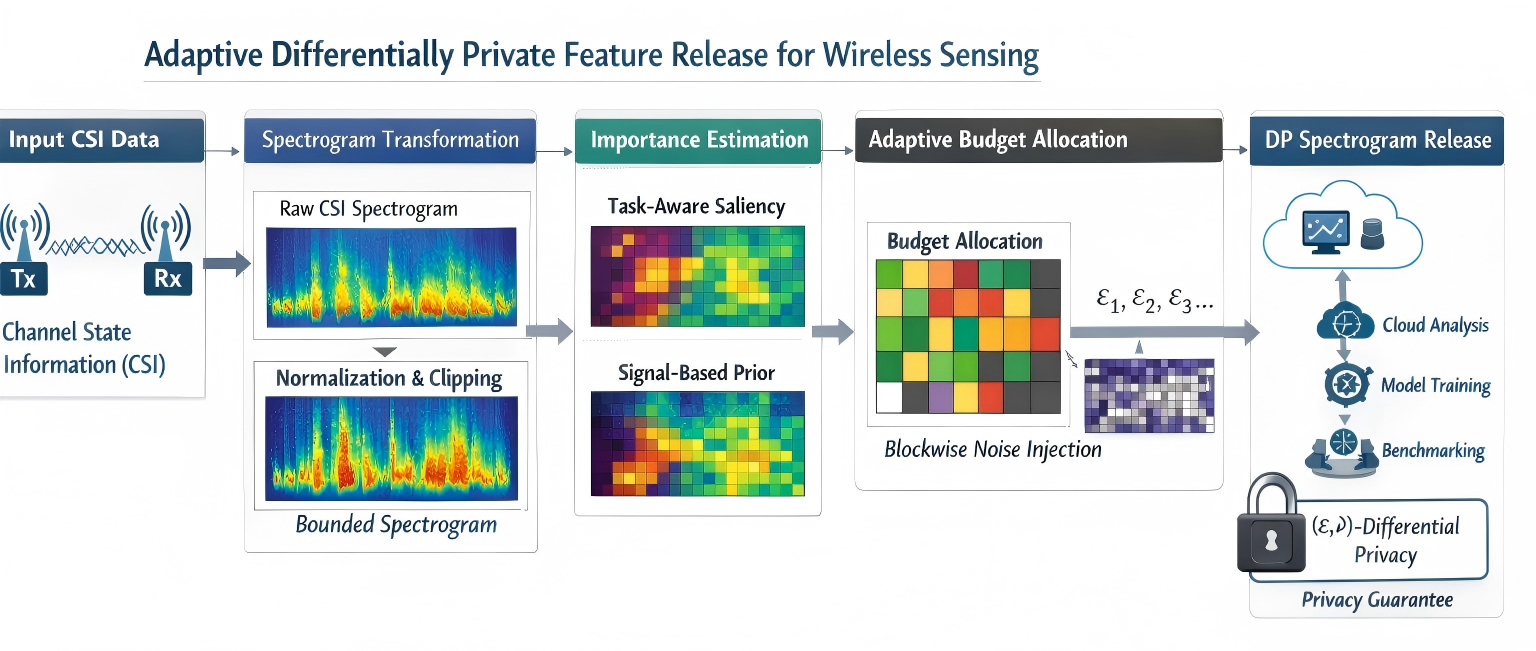}
  \caption{System overview}
  \label{fig:pipeline}
\end{figure*}

We study \emph{differentially private (DP) feature release} for wireless sensing, focusing on CSI-derived spectrogram features that are commonly used by modern deep models for human activity recognition, respiration monitoring, and related tasks. Our goal is to release a representation that remains useful for downstream inference while providing a formal privacy guarantee against inference attacks on the released features. The key technical idea is that CSI spectrograms are highly non-uniform: task-relevant information concentrates in specific time--frequency regions (e.g., Doppler bands for motion or narrow periodic components for respiration), and the concentration pattern depends on the task and environment. Therefore, instead of applying a uniform DP noise level across all bins, we allocate the privacy budget \emph{adaptively} according to an importance map, concentrating privacy spending where it yields the largest utility under a fixed global privacy budget.

\subsection{Problem setting and release protocol}
\label{subsec:method_setting}

Let $\mathcal{D}$ denote a dataset of CSI windows. Each window corresponds to a short time segment (e.g., a few hundred milliseconds to a few seconds) that will be converted into a 2D time--frequency feature map. We consider a feature-release setting in which a data owner (or edge device) releases a DP-protected feature $\widetilde{\mathbf{S}}$ to an untrusted consumer (e.g., cloud trainer, third-party evaluator, or analytics service). The consumer may train or run downstream models on $\widetilde{\mathbf{S}}$, but should not be able to infer sensitive information about individual windows (sample-level privacy) beyond the DP guarantee.

\paragraph{Neighboring datasets.}
Our primary guarantee is \emph{sample-level} DP: two neighboring datasets $\mathcal{D}$ and $\mathcal{D}'$ differ in exactly one CSI window. This is a practical choice when windows are independently released or when a device periodically uploads window-level features. The same machinery can be extended to user-level DP by grouping all windows from one person and applying group privacy or per-user accounting; we discuss this extension in Section.

\paragraph{Overview of our pipeline.}
As illustrated in Fig.~\ref{fig:pipeline}, the release pipeline comprises: (i) CSI-to-spectrogram transformation and normalization; (ii) sensitivity control via clipping; (iii) importance estimation on the spectrogram (task-aware or task-agnostic); (iv) adaptive privacy-budget allocation across time--frequency blocks; (v) blockwise Gaussian mechanism for DP noise injection; and (vi) privacy accounting under composition across blocks and across repeated releases.

\subsection{CSI spectrogram construction and normalization}
\label{subsec:method_spectrogram}

We begin with CSI extracted from commodity Wi-Fi devices. For a given window, we form a complex CSI sequence $\mathbf{h}(t)\in\mathbb{C}^{N_c}$, where $N_c$ is the number of subcarriers (or an equivalent channel dimension). We compute a time--frequency representation using a short-time Fourier transform (STFT) along the temporal axis for each subcarrier/channel and then aggregate across channels to obtain a single spectrogram-like map. Concretely, for a window indexed by $i$, we define the (magnitude) spectrogram as
\begin{equation}
\mathbf{S}_i(\tau,\omega) \;=\; \frac{1}{N_c}\sum_{c=1}^{N_c}\left|\sum_{t=0}^{T-1} w(t-\tau)\, \mathbf{h}_{i,c}(t)\, e^{-j\omega t}\right| ,
\tag{1}
\end{equation}
\noindent{\small where $\mathbf{h}_{i,c}(t)$ is the complex CSI on subcarrier $c$ at time $t$ for window $i$, $w(\cdot)$ is the analysis window, $\tau$ indexes time frames, $\omega$ indexes frequency bins, and $T$ is the number of samples in the window.}

Equation~(1) yields a nonnegative matrix $\mathbf{S}_i\in\mathbb{R}_{\ge 0}^{T_f\times F}$ with $T_f$ frames and $F$ frequency bins. To stabilize sensitivity and facilitate DP noise calibration, we apply a bounded normalization (e.g., log-compression followed by scaling to $[0,1]$):
\begin{equation}
\mathbf{X}_i \;=\; \mathrm{Norm}\!\left(\log(1+\mathbf{S}_i)\right),
\qquad \mathbf{X}_i\in[0,1]^{T_f\times F}.
\tag{2}
\end{equation}
\noindent{\small where $\mathrm{Norm}(\cdot)$ denotes an elementwise affine normalization (e.g., min--max or robust percentile scaling) that maps values into $[0,1]$, and $\mathbf{X}_i$ is the released pre-noise feature.}

The bounded range in (2) is not merely a modeling convenience: it is central to DP because it allows us to establish a finite sensitivity bound for the released feature.

\subsection{Sensitivity control via clipping}
\label{subsec:method_clipping}

In practice, even after normalization, outliers may occur due to hardware artifacts, abrupt interference, or sporadic channel events. We further apply clipping to ensure that each window feature has bounded $\ell_2$ norm (or bounded per-entry range), which yields a clean sensitivity bound for the subsequent Gaussian mechanism. We use matrix-vectorization for simplicity: let $\mathbf{x}_i=\mathrm{vec}(\mathbf{X}_i)\in\mathbb{R}^{d}$, where $d=T_f\cdot F$. We apply $\ell_2$ clipping:
\begin{equation}
\bar{\mathbf{x}}_i \;=\; \mathbf{x}_i\cdot \min\!\left(1,\frac{C}{\|\mathbf{x}_i\|_2}\right),
\tag{3}
\end{equation}
\noindent{\small where $C>0$ is the clipping threshold, $\|\cdot\|_2$ is the Euclidean norm, $\mathbf{x}_i$ is the vectorized normalized spectrogram, and $\bar{\mathbf{x}}_i$ is the clipped representation.}

We then define the released (pre-noise) matrix $\bar{\mathbf{X}}_i=\mathrm{unvec}(\bar{\mathbf{x}}_i)\in\mathbb{R}^{T_f\times F}$. Under sample-level adjacency, changing one window changes at most one clipped vector. Therefore, the $\ell_2$ sensitivity of the identity query that outputs $\bar{\mathbf{x}}_i$ is bounded by
\begin{equation}
\Delta_2 \;\triangleq\; \max_{\mathcal{D}\sim\mathcal{D}'} \left\|\bar{\mathbf{x}}(\mathcal{D})-\bar{\mathbf{x}}(\mathcal{D}')\right\|_2 \;\le\; 2C,
\tag{4}
\end{equation}
\noindent{\small where $\mathcal{D}\sim\mathcal{D}'$ denotes neighboring datasets, $\bar{\mathbf{x}}(\mathcal{D})$ denotes the released clipped vector for the differing record, and $C$ is the clipping threshold in (3).}

The bound in (4) is standard: since both $\bar{\mathbf{x}}$ and $\bar{\mathbf{x}}'$ have norm at most $C$, their difference has norm at most $2C$. This bound is conservative but leads to a simple and robust DP calibration.

\subsection{Importance estimation on the time--frequency plane}
\label{subsec:method_importance}

To allocate privacy budget adaptively, we estimate an importance map that captures how much each time--frequency region contributes to downstream utility. We support two design modes depending on whether a downstream task model is available:

\paragraph{Task-aware (gradient-based) importance.}
When a surrogate downstream model $f_\theta$ is available (e.g., a HAR classifier trained on private data at the data owner), we measure importance via the gradient magnitude of a task loss $\mathcal{L}$ with respect to the input feature. For window $i$ with label $y_i$, we define
\begin{equation}
\mathbf{G}_i \;=\; \left|\nabla_{\bar{\mathbf{X}}_i}\, \mathcal{L}\!\left(f_\theta(\bar{\mathbf{X}}_i),\, y_i\right)\right|,
\qquad
\mathbf{W}_i \;=\; \frac{\mathbf{G}_i}{\sum_{\tau,\omega} \mathbf{G}_i(\tau,\omega) + \epsilon_0},
\tag{5}
\end{equation}
\noindent{\small where $\mathcal{L}$ is a task loss (e.g., cross-entropy), $f_\theta$ is a surrogate model, $\mathbf{G}_i$ is the elementwise gradient magnitude map, $\mathbf{W}_i$ is the normalized importance map, and $\epsilon_0>0$ avoids division by zero.}

Intuitively, bins that strongly influence the loss (large gradient magnitude) are more important for the task, and thus should receive a larger privacy budget (i.e., less noise) to preserve utility. Importantly, the importance map computation is performed \emph{locally} by the data owner prior to release; the map itself is not released and can be treated as internal metadata. If desired, one can also apply DP to $\mathbf{W}_i$, but we found it sufficient in practice to keep it on-device and only release the DP-protected feature.

\paragraph{Task-agnostic (signal-prior) importance.}
If no task model is available, we can construct $\mathbf{W}_i$ from signal priors such as local energy concentration, spectral entropy, or band-specific heuristics (e.g., focusing around known respiration bands). The remainder of the pipeline is unchanged; only the construction of $\mathbf{W}_i$ differs.

\subsection{Adaptive privacy budget allocation}
\label{subsec:method_budget}

We allocate privacy budgets across the time--frequency plane under a global privacy constraint. To reduce accountant complexity and improve stability, we partition $\bar{\mathbf{X}}_i$ into $B$ non-overlapping blocks (rectangles) $\{\mathcal{B}_b\}_{b=1}^B$, and enforce a uniform budget within each block. Let $w_{i,b}=\sum_{(\tau,\omega)\in\mathcal{B}_b}\mathbf{W}_i(\tau,\omega)$ be the block importance mass. We then map $w_{i,b}$ to a per-block privacy budget $\varepsilon_{i,b}$ using a monotone allocation rule with lower/upper bounds:
\begin{equation}
\begin{split}
\varepsilon_{i,b} &= \varepsilon_{\min} + (\varepsilon_{\max}-\varepsilon_{\min})\cdot \frac{w_{i,b}^{\gamma}}{\sum_{b'=1}^{B} w_{i,b'}^{\gamma}}, \\
\sum_{b=1}^{B} \varepsilon_{i,b} &= \varepsilon_{\min}B + (\varepsilon_{\max}-\varepsilon_{\min}).
\end{split}
\tag{6}
\end{equation}
\noindent{\small where $w_{i,b}$ is the importance of block $b$, $\gamma\ge 1$ controls allocation sharpness, $\varepsilon_{\min}$ and $\varepsilon_{\max}$ set per-block budget bounds, and $\varepsilon_{i,b}$ is the allocated privacy budget for block $b$ in window $i$.}

Equation~(6) is a practical budget scheduler: (i) it is monotone in $w_{i,b}$; (ii) it avoids starving low-importance blocks via $\varepsilon_{\min}$; and (iii) it prevents concentrating almost all budget in a tiny region via $\varepsilon_{\max}$. The exponent $\gamma$ allows the mechanism to smoothly interpolate between near-uniform allocation ($\gamma\approx 1$) and highly concentrated spending ($\gamma\gg 1$). While (6) does not directly enforce a single scalar $\varepsilon$ for the whole release, it provides a convenient \emph{budget vector} that can be paired with a composition accountant to compute the final global $(\varepsilon,\delta)$ guarantee. In our implementation, we treat $\{\varepsilon_{i,b}\}$ as the mechanism parameters and compute the overall privacy loss via Rényi DP accounting under composition across blocks and repeated releases (see below). This separation (allocation first, accounting second) yields a modular design that is easy to deploy and audit.

\subsection{DP feature release via blockwise Gaussian mechanism}
\label{subsec:method_gaussian}

Given per-block budgets, we apply a blockwise Gaussian mechanism to $\bar{\mathbf{x}}_i$ (or equivalently $\bar{\mathbf{X}}_i$). Let $\bar{\mathbf{x}}_{i,b}\in\mathbb{R}^{d_b}$ denote the vector of entries in block $\mathcal{B}_b$, where $d_b$ is the block size. For each block we release
\begin{equation}
\widetilde{\mathbf{x}}_{i,b} \;=\; \bar{\mathbf{x}}_{i,b} \;+\; \mathbf{z}_{i,b},
\qquad
\mathbf{z}_{i,b} \sim \mathcal{N}\!\left(\mathbf{0},\, \sigma_{i,b}^2 \mathbf{I}_{d_b}\right),
\tag{7}
\end{equation}
\noindent{\small where $\widetilde{\mathbf{x}}_{i,b}$ is the DP-protected released block, $\mathbf{z}_{i,b}$ is Gaussian noise, $\sigma_{i,b}$ is the noise standard deviation for block $b$, and $\mathbf{I}_{d_b}$ is the $d_b\times d_b$ identity matrix.}

To connect budgets to noise scales, we use the standard Gaussian calibration with sensitivity controlled by clipping. For sample-level DP, the $\ell_2$ sensitivity of the block query is at most $2C$ (by (4)), and we set
\begin{equation}
\sigma_{i,b} \;=\; \kappa \cdot \frac{2C}{\varepsilon_{i,b}},
\tag{8}
\end{equation}
\noindent{\small where $C$ is the clipping threshold, $\varepsilon_{i,b}$ is the allocated block budget from (6), and $\kappa$ is a constant determined by the chosen accountant and target $\delta$ (e.g., via standard Gaussian DP bounds or via RDP conversion).}

In practice, rather than directly using a fixed $\kappa$ from a loose bound, we implement a Rényi DP (RDP) accountant to obtain tighter privacy accounting under composition. The mapping in (8) then serves as an initialization, and we compute the final global privacy parameters via accountant conversion.

\subsection{Privacy accounting under composition}
\label{subsec:method_accounting}

Feature release is often repeated over time (multiple windows per user/device), and within each window we release multiple blocks. Therefore, we must account for composition both \emph{within-window} (across blocks) and \emph{across-windows} (across time). We adopt an RDP-based accountant because it typically yields tighter bounds than basic or advanced composition, especially for Gaussian mechanisms.

Let $\varepsilon^{\mathrm{RDP}}(\alpha; \sigma)$ denote the RDP of order $\alpha>1$ for a Gaussian mechanism with noise scale $\sigma$ and sensitivity $\Delta_2$. For each block mechanism in (7), we can compute an RDP contribution $\varepsilon_{i,b}^{\mathrm{RDP}}(\alpha)$ (closed-form for Gaussian). Under composition, RDP adds linearly. For a sequence of releases $\mathcal{M}_1,\ldots,\mathcal{M}_K$ (spanning blocks and windows), the total RDP is
\begin{equation}
\varepsilon_{\mathrm{tot}}^{\mathrm{RDP}}(\alpha) \;=\; \sum_{k=1}^{K} \varepsilon_{k}^{\mathrm{RDP}}(\alpha),
\tag{9}
\end{equation}
\noindent{\small where $\alpha$ is the Rényi order, $K$ is the number of composed mechanisms (blocks across windows), and $\varepsilon_{k}^{\mathrm{RDP}}(\alpha)$ is the RDP contribution of the $k$-th mechanism.}

Finally, we convert the composed RDP bound into an $(\varepsilon,\delta)$-DP guarantee using the standard RDP-to-DP conversion:
\begin{equation}
\varepsilon(\delta) \;=\; \min_{\alpha>1}\left\{\varepsilon_{\mathrm{tot}}^{\mathrm{RDP}}(\alpha) \;+\; \frac{\log(1/\delta)}{\alpha-1}\right\}.
\tag{10}
\end{equation}
\noindent{\small where $\delta\in(0,1)$ is the target failure probability, $\alpha$ is optimized over a finite grid, and $\varepsilon(\delta)$ is the resulting DP parameter for the entire release process.}

Equations (9)--(10) provide a transparent audit trail: given the sequence of released blocks and their noise scales, we can compute the cumulative privacy loss and enforce a global budget cap. In deployment, this allows a device to stop releasing features once it exhausts its privacy budget or to dynamically adapt $\varepsilon_{i,b}$ over time depending on the remaining budget.

\subsection{Implementation details and practical considerations}
\label{subsec:method_practical}

\paragraph{Block granularity.}
Block partitioning trades off adaptivity and stability. Very small blocks (near per-bin) allow fine-grained budget allocation but increase composition cost and may amplify variance in importance estimation. Larger blocks reduce accountant overhead and smooth allocation but may waste budget if important and unimportant bins coexist in one block. In our experiments, we choose rectangular blocks aligned with the STFT grid (e.g., grouping a few consecutive time frames and frequency bins) to respect local correlation in spectrograms.

\paragraph{Importance stability and leakage.}
Although $\mathbf{W}_i$ is not released, it influences the release via $\varepsilon_{i,b}$. In principle, this can create a side channel if an adversary observes variations in noise across regions. We mitigate this by (i) enforcing bounded allocation via $(\varepsilon_{\min},\varepsilon_{\max})$, (ii) using fixed block partitions, and (iii) optionally using a \emph{public} (task-agnostic) importance prior for deployments where labels or private models are unavailable. Empirically, bounded allocations preserve the primary benefit (protecting low-importance regions with stronger noise while preserving high-importance regions) while preventing extreme heterogeneity.

\paragraph{Streaming release.}
For streaming operation, the global accountant maintains the cumulative privacy loss per device (or per user, if user-level DP is targeted). Once the budget is exhausted, the device can (i) stop releasing; (ii) increase noise by shrinking $\varepsilon_{i,b}$; or (iii) downsample in time to reduce $K$ in (9). This control loop is particularly relevant under evolving Wi-Fi MAC behavior, where measurement availability may fluctuate.

\paragraph{Compatibility with downstream training.}
The released feature $\widetilde{\mathbf{X}}_i$ can be consumed by any downstream model without modification because DP is preserved under post-processing. In practice, training with DP-noised inputs benefits from standard regularization and augmentation; however, our contribution is orthogonal to DP-SGD, since we protect \emph{released features} rather than private gradients during centralized training.

Overall, this methodology provides an auditable, modular DP feature-release mechanism tailored to CSI spectrograms. It leverages the non-uniform structure of wireless sensing representations through importance-aware budget allocation, while maintaining formal privacy guarantees under rigorous composition accounting.

\section{Experimental Setup}
\label{sec:exp}

This section describes the datasets, hardware/software environment, preprocessing pipeline, evaluation protocol, and metrics used to assess the proposed \emph{differentially private wireless-sensing feature release} framework. Our goal is to quantify (i) \textbf{utility} for downstream sensing tasks when only privatized features are available to a data consumer, and (ii) \textbf{privacy} against realistic inference attacks that exploit released CSI-derived representations. We design experiments to cover both \emph{activity-centric} and \emph{fine-grained} human sensing scenarios, and to stress-test robustness across domains (users, environments, and WiFi bands).

\subsection{Datasets and Downstream Tasks}
\label{subsec:datasets}

We evaluate on two public WiFi sensing benchmarks and one controlled in-house dataset. The public datasets provide reproducibility and diverse sensing conditions; the in-house dataset reflects a typical \emph{feature publishing} scenario where raw CSI cannot be shared due to privacy or policy constraints. Table~\ref{tab:datasets} summarizes the benchmarks and their key statistics.

\paragraph{WiMANS: multi-user activity sensing with identity/location attributes.}
WiMANS is a benchmark dataset for WiFi-based multi-user sensing that contains dual-band CSI (2.4\,GHz and 5\,GHz) and synchronized videos \cite{huang2024wimans}. Each sample is a 3-second recording with up to 5 concurrent users, annotated with per-user \emph{identity}, \emph{location}, and \emph{activity}. WiMANS includes 6 subjects and 9 daily activities (Nothing, Walking, Rotation, Jumping, Waving, Lying Down, Picking Up, Sitting Down, Standing Up) collected in three indoor environments (classroom, meeting room, empty room) \cite{huang2024wimans}. In WiMANS, each CSI sample is recorded at a high packet rate and has a consistent temporal length (nominally 3000 time steps per 3 seconds, with mild packet loss depending on band) \cite{huang2024wimans}. \textbf{Utility task:} we focus on multi-user human activity recognition (MU-HAR) using the activity labels as the downstream objective. \textbf{Privacy task:} because WiMANS provides anonymized identity and location annotations, it naturally supports privacy-leakage evaluation via (i) identity inference and (ii) location inference from the released features.

\paragraph{Person-in-WiFi 3D: multi-person 3D pose estimation.}
Person-in-WiFi 3D is an end-to-end multi-person 3D pose estimation benchmark with WiFi signals \cite{yan2024person}. It provides over 97K synchronized samples collected in a 4\,m$\times$3.5\,m area, with seven volunteers and up to three people in the scene \cite{yan2024person}. The collection uses one transmitter and three receivers (Intel 5300 NICs), operating on a 5\,GHz channel, and synchronizes CSI with an Azure Kinect stream for pose supervision \cite{yan2024person}. \textbf{Utility task:} we treat 3D pose estimation as a fine-grained regression benchmark to test whether the released DP features preserve subtle motion cues beyond coarse activities.

\paragraph{Resp-CSI (in-house): respiration waveform and rate estimation under interference.}
To reflect healthcare-oriented sensing where data sharing is particularly sensitive, we collect a small in-house dataset for contactless respiration monitoring inspired by commodity-WiFi respiratory sensing systems \cite{zhao2024wi,xie2024robust}. Resp-CSI contains 24 participants, each recorded in two rooms (bedroom-like and office-like) under three conditions: (i) normal breathing, (ii) paced deep breathing, and (iii) breathing with an interfering bystander moving slowly within the sensing area. Each trial lasts 120\,s and is repeated twice, resulting in 24$\times$2$\times$3$\times$2 = 288 trials. Ground-truth respiration is obtained using a chest belt sensor (sampled at 50\,Hz), synchronized with CSI time stamps. \textbf{Utility task:} we evaluate respiration \emph{rate} estimation (breaths per minute) and respiration \emph{waveform} tracking. \textbf{Privacy task:} we evaluate membership inference and attribute inference (room/condition) on the released representations, which is a realistic threat when a third party can train a classifier over published features.
\begin{table*}[t]
\centering
\small

\setlength{\tabcolsep}{4pt} 

\newcolumntype{Y}{>{\centering\arraybackslash}X}

\begin{tabularx}{\textwidth}{l >{\raggedright\arraybackslash}p{2.8cm} Y c c Y Y}
\toprule
Dataset & Modality & Environments & Subjects & Samples & Labels (utility) & Labels (privacy) \\
\midrule
WiMANS \cite{huang2024wimans} & 
CSI (2.4/5\,GHz) + video & 
3 rooms; 5 locations & 
6 & 
11,286 & 
9 activities (multi-user) & 
6 identities + 5 locations \\ 
\addlinespace[4pt] 

Person-in-WiFi 3D \cite{yan2024person} & 
CSI (5\,GHz) + RGB-D & 
1 area (4$\times$3.5\,m) & 
7 & 
$\approx$97,000 frames & 
3D poses (14 joints/person) & 
subject ID; scenario (1/2/3 persons) \\ 
\addlinespace[4pt]

Resp-CSI (in-house) & 
CSI (5\,GHz) + belt sensor & 
2 rooms & 
24 & 
288 trials & 
resp. rate + waveform & 
room/condition; membership \\
\bottomrule
\end{tabularx}
\caption{Datasets and tasks used in evaluation. WiMANS supports both utility (HAR) and privacy analysis due to its rich annotations.}
\label{tab:datasets}
\end{table*}

\subsection{Hardware, Data Acquisition, and Compute Platform}
\label{subsec:hardware}

\paragraph{Public-dataset acquisition setups.}
For WiMANS, the data are collected using two off-the-shelf computers equipped with Intel 5300 NICs and the Linux 802.11n CSI tool \cite{huang2024wimans}. The transmitter/receiver each has 3 antennas, and CSI is measured on 30 subcarriers, yielding a per-time-step CSI tensor of size $3\times 3\times 30$ (Tx antennas $\times$ Rx antennas $\times$ subcarriers) \cite{huang2024wimans}. Each 3-second sample is obtained by sending 3000 packets at 1000 packets per second, producing a nominal shape of $3000\times 3\times 3\times 30$, with packet loss rates depending on band \cite{huang2024wimans}. WiMANS provides CSI for both 2.4\,GHz (channel 12) and 5\,GHz (channel 64), enabling us to test band-specific privacy/utility behavior \cite{huang2024wimans}.

For Person-in-WiFi 3D, the dataset uses four ThinkPad X201 laptops (one transmitter and three receivers) equipped with Intel 5300 cards \cite{yan2024person}. The transmitter operates on a 5\,GHz channel and sends packets at 300 packets per second; the system synchronizes CSI with Azure Kinect RGB-D frames at 15 fps \cite{yan2024person}. Each CSI sample aligned to a video frame is stored as a tensor of shape $1\times 3\times 3\times 30\times 20$ corresponding to (\#Tx, \#Rx, \#antenna, \#subcarrier, \#time) \cite{yan2024person}.

\paragraph{Resp-CSI acquisition setup (in-house).}
We build a commodity WiFi sensing link using a single transmitter and a single receiver, each equipped with an Intel 5300 NIC (3 antennas enabled). The transmitter broadcasts on a fixed 5\,GHz channel (20\,MHz bandwidth) at 800 packets per second to ensure sufficient temporal resolution for respiration-induced micro-motions; the receiver runs in monitor mode and logs CSI using the same CSI toolchain as commodity CSI sensing pipelines. The two devices are placed 3.0\,m apart at a height of 1.0\,m. Participants sit 1.5--2.0\,m from the line-of-sight (LoS) path and breathe naturally or following a metronome. For the interference condition, an additional person walks slowly along a 2\,m arc behind the participant at $\approx$0.2\,m/s to emulate the ``interfering individual'' setting discussed in robust respiration sensing \cite{xie2024robust}. A chest belt sensor provides the respiration waveform reference at 50\,Hz; belt samples are time-aligned to CSI via NTP-synchronized timestamps.

\paragraph{Training and inference compute.}
All models are trained in PyTorch (v2.2) with CUDA 12.x. Unless otherwise stated, training is performed on a workstation with an NVIDIA RTX 4090 (24\,GB), Intel i9-class CPU, and 64\,GB RAM. We report the mean and standard deviation over three random seeds for all main experiments. For deployment-related measurements (feature-size and runtime), we additionally benchmark on an edge-like mini-PC (Intel i7-1260P, 16\,GB RAM) to estimate the cost of on-device feature extraction and DP perturbation.

\begin{table}[t]
\centering
\small
\setlength{\tabcolsep}{5pt}
\begin{tabular}{lcc}
\toprule
Component & Setting & Notes \\
\midrule
GPU & RTX 4090 (24\,GB) & training + attack models \\
CPU & Intel i9-class & data loading + DP noise \\
RAM & 64\,GB & caching spectrograms \\
OS & Ubuntu 22.04 LTS & reproducible environment \\
Framework & PyTorch 2.2 + CUDA 12.x & AMP enabled \\
\bottomrule
\end{tabular}
\caption{Compute platform used for training and evaluation.}
\label{tab:compute}
\end{table}

\subsection{Preprocessing and CSI-to-Spectrogram Construction}
\label{subsec:preprocess}

We standardize preprocessing across datasets to isolate the effect of privacy mechanisms. Starting from complex CSI $H(t,f)$ (time index $t$, subcarrier index $f$), we extract amplitude and optionally phase (with denoising when available). Following the common observation that CSI phase can be noisy due to hardware offsets, we use phase only when the dataset provides a denoised phase (Person-in-WiFi 3D includes a phase denoising step based on linear transformation \cite{yan2024person}). For WiMANS and Resp-CSI, we mainly use amplitude and apply a light temporal smoothing to reduce packet-level jitter.

\paragraph{Packet-loss handling.}
WiMANS reports non-negligible packet loss, especially in 2.4\,GHz \cite{huang2024wimans}. We therefore resample each antenna-pair/subcarrier stream to a fixed length per window using linear interpolation on timestamps. This makes the subsequent STFT consistent across samples and prevents DP clipping from being biased by missing values.

\paragraph{Windowing strategy.}
For WiMANS, each sample is 3\,s; we keep the native 3\,s window as the unit of release, matching its annotation granularity \cite{huang2024wimans}. For Resp-CSI, we segment each 120\,s trial into non-overlapping 6\,s windows (20 windows per trial), which is long enough to cover multiple breathing cycles while still supporting DP composition accounting. For Person-in-WiFi 3D, we stack consecutive CSI tensors aligned to video frames to form 2\,s clips (30 frames at 15\,fps), which provides sufficient temporal context for stable pose estimation.

\paragraph{Spectrogram extraction.}
We compute a short-time Fourier transform (STFT) over the time axis for each subcarrier stream and antenna pair. Concretely, for a real-valued amplitude sequence $x[n]$ in a window, we compute:
\begin{equation}
S[m,k] = \sum_{n=0}^{N-1} x[n+mH]\, w[n]\, e^{-j 2\pi kn/N},
\label{eq:stft}
\end{equation}
where $w[n]$ is a Hann window, $N$ is the FFT size, and $H$ is the hop length. For WiMANS and Resp-CSI, we set $N=256$ and $H=64$ (75\% overlap), producing stable time-frequency resolution for typical human motions. We take log-magnitude $\log(|S|+\epsilon_0)$ with $\epsilon_0=10^{-6}$ and then apply per-frequency normalization using training-set statistics. For Person-in-WiFi 3D, since the native CSI tensor already aggregates a short time slice per frame \cite{yan2024person}, we first concatenate the time dimension across frames and then apply STFT.

\paragraph{Representation shape.}
To keep the released feature size manageable, we compress along antenna-pair and subcarrier dimensions by a learnable front-end (defined in the methodology section) and release a compact spectrogram embedding of dimension $d=256$ per window by default. In ablations, we also test $d\in\{128,512\}$ to study privacy/utility sensitivity to representation capacity.

\subsection{Evaluation Protocol: Splits and Generalization Settings}
\label{subsec:splits}

We evaluate both \emph{in-distribution} performance and \emph{cross-domain} generalization because privacy-preserving release often faces distribution shift when a consumer trains on one environment and deploys in another.

\paragraph{WiMANS splits.}
WiMANS organizes samples into user groups with fixed numbers of users and environments \cite{huang2024wimans}. To avoid leakage due to correlated sampling within a group, we split by \emph{group index}: 70\% groups for training, 10\% for validation, and 20\% for testing, stratified by environment and number of users (0--5). We report results separately on 2.4\,GHz and 5\,GHz, and also on a dual-band setting where both bands are processed and fused at the feature level. Additionally, we report:
(i) \textbf{Leave-one-environment-out (LOEO):} train on two environments and test on the third; and
(ii) \textbf{Leave-two-identities-out (LTIO):} train on 4 identities and test on the remaining 2 identities. The LOEO setting mirrors the practical deployment challenge where a consumer receives privatized features collected in unseen rooms.

\paragraph{Person-in-WiFi 3D splits.}
We follow the dataset's standard cross-domain evaluations \cite{yan2024person}. Specifically, we report:
(i) \textbf{Leave-one-person-out (LOPO):} train on six volunteers and test on the held-out volunteer; and
(ii) \textbf{Scenario generalization:} train on 1-person and 2-person scenes and test on 3-person scenes to quantify robustness under increased multi-user interference. These settings are particularly relevant for privacy because multi-person scenes may amplify identity leakage due to distinctive motion signatures.

\paragraph{Resp-CSI splits.}
We use subject-disjoint splits: 16 subjects for training, 4 for validation, and 4 for testing. We also evaluate an interference generalization setting where models are trained on (normal + paced) conditions and tested on the interfering-bystander condition, following the motivation of interference-robust respiration sensing \cite{xie2024robust}.

\subsection{Compared Methods and Baselines}
\label{subsec:baselines}

We compare the proposed \emph{privacy-budget adaptive} feature release against both non-private and privacy-agnostic alternatives. All approaches release the same embedding dimension $d$ unless stated otherwise, so improvements reflect privacy allocation rather than bandwidth differences.

\paragraph{Non-private feature release (No-DP).}
We release the learned spectrogram embedding without adding DP noise. This upper-bounds utility but provides no formal privacy protection.

\paragraph{Uniform-DP (fixed budget).}
We allocate a fixed privacy budget per window, using the same $\epsilon$ across all time-frequency regions. This baseline tests whether adaptivity is necessary.

\paragraph{Heuristic importance-weighted DP.}
We allocate more budget to high-energy or high-saliency regions using a deterministic rule (e.g., based on spectrogram magnitude percentiles) rather than a learned policy. This tests whether simple heuristics suffice.

\paragraph{Random allocation.}
We randomly allocate the same total budget across frequency bands. This controls for the possibility that any non-uniformity improves results.

\paragraph{Adversarial perturbation robustness (stress test).}
Wireless sensing pipelines can be attacked via perturbations at the packet level, which can affect the feature distribution and downstream inference \cite{cao2024security,li2024practical}. While our main focus is privacy leakage from released features, we additionally report utility degradation when the released features are computed from mildly perturbed CSI streams (e.g., random packet dropping or small additive noise on amplitude) to confirm that DP noise does not catastrophically amplify adversarial fragility.

\subsection{Utility Metrics}
\label{subsec:utility-metrics}

We choose task-appropriate utility metrics that reflect both overall correctness and class balance.

\paragraph{Multi-user HAR on WiMANS.}
WiMANS samples can include multiple simultaneous activities (one per user) \cite{huang2024wimans}. We treat MU-HAR as a multi-label classification problem over the 9 activity classes, and report:
(i) micro-averaged accuracy, (ii) macro-F1, and (iii) per-class F1 to diagnose which activities are most sensitive to privacy noise. For completeness, we also report the single-label HAR accuracy on the 1-user subset (Nu=1) to align with prior single-user sensing literature.

\paragraph{3D pose estimation on Person-in-WiFi 3D.}
We use mean per-joint position error (MPJPE, in mm), the standard metric in 3D pose estimation \cite{yan2024person}. MPJPE is computed after matching predicted persons to ground-truth persons via Hungarian assignment (set prediction), consistent with multi-person evaluation protocols.

\paragraph{Respiration sensing on Resp-CSI.}
We report (i) mean absolute error (MAE) of respiration rate (breaths/min) and (ii) Pearson correlation between predicted and ground-truth respiration waveforms. For waveform correlation, both signals are band-pass filtered (0.1--0.7\,Hz) and normalized per window.

\begin{table*}[t]
\centering
\small
\setlength{\tabcolsep}{6pt}
\begin{tabular}{lll}
\toprule
Task & Metric & Definition \\
\midrule
WiMANS MU-HAR & Accuracy & fraction of correctly predicted activity labels \\
WiMANS MU-HAR & Macro-F1 & average F1 across 9 activities (class-balanced) \\
Pose (3D) & MPJPE (mm) & mean Euclidean error per joint after matching \\
Respiration & Rate MAE & MAE between predicted and true breaths/min \\
Respiration & Waveform corr. & Pearson correlation over each 6\,s window \\
\bottomrule
\end{tabular}
\caption{Utility metrics used for downstream sensing tasks.}
\label{tab:utility-metrics}
\end{table*}

\subsection{Privacy Metrics and Attack Models}
\label{subsec:privacy-metrics}

We evaluate privacy using both \emph{formal DP parameters} (reported $\epsilon,\delta$ under a consistent adjacency definition) and \emph{empirical leakage} under standard inference attacks. This dual view is important: DP offers worst-case guarantees, while empirical attacks reveal which attributes remain practically inferable from released embeddings.

\paragraph{Membership inference attack (MIA).}
The attacker observes released features and tries to decide whether a specific window (or a subject) was included in the producer's training set. We adopt a shadow-model approach: the attacker trains $M$ shadow models (we use $M=4$) on disjoint subsets, collects confidence vectors or intermediate embeddings, and trains a binary attack classifier. We report:
(i) ROC-AUC, (ii) attack accuracy at the optimal threshold, and (iii) attack advantage (TPR--FPR at the best threshold). We perform both record-level MIA (window membership) and subject-level MIA (all windows of a person).

\paragraph{Identity and location inference.}
On WiMANS, identity and location labels are available in addition to activity \cite{huang2024wimans}. We treat identity inference as an attacker training a classifier to predict the user identity from released features. Similarly, we train a location inference classifier for the five locations (A--E) \cite{huang2024wimans}. These attacks reflect privacy threats where anonymized published features can be re-identified by linking to auxiliary data. We report Top-1 accuracy and macro-F1 for both attacks.

\paragraph{Attribute inference on Resp-CSI.}
For Resp-CSI, we define two sensitive attributes: \emph{room type} (bedroom-like vs office-like) and \emph{condition} (normal, paced, interference). We train multi-class classifiers on the released features and report Top-1 accuracy. This measures whether private contextual information can be inferred beyond the intended health signal.

\paragraph{Reconstruction attack (optional).}
To quantify how much raw spectrogram structure leaks through the embedding, we optionally train a decoder that reconstructs a spectrogram from the released embedding. We report reconstruction MSE and SSIM. While reconstruction is not always the attacker’s objective, it provides an interpretable proxy of information leakage.

\begin{table*}[t]
\centering
\small
\setlength{\tabcolsep}{6pt}
\begin{tabular}{lll}
\toprule
Privacy aspect & Attack & Reported metric(s) \\
\midrule
Membership & window-/subject-level MIA & AUC, accuracy, advantage \\
Re-identification & identity inference (WiMANS) & Top-1 acc., macro-F1 \\
Context leakage & location inference (WiMANS) & Top-1 acc., macro-F1 \\
Context leakage & room/condition inference (Resp-CSI) & Top-1 acc. \\
Reconstruction & spectrogram decoder & MSE, SSIM (optional) \\
\bottomrule
\end{tabular}
\caption{Privacy evaluation: attack models and leakage metrics.}
\label{tab:privacy-metrics}
\end{table*}

\subsection{DP Configuration and Hyperparameters}
\label{subsec:dp-config}

We evaluate a range of privacy budgets typical for practical deployments, and we keep $\delta$ fixed across all experiments for comparability. Unless otherwise stated, we set $\delta = 10^{-5}$ for WiMANS and Person-in-WiFi 3D, and $\delta = 10^{-6}$ for Resp-CSI due to the larger number of windows per subject.

\paragraph{Budgets.}
We sweep $\epsilon \in \{0.25, 0.5, 1, 2, 4, 8\}$ to obtain privacy--utility curves. For each $\epsilon$, we apply the same total budget per released window; adaptive methods redistribute the budget internally across time-frequency regions but preserve the same global DP accounting.

\paragraph{Sensitivity control and clipping.}
We set the per-window $\ell_2$ clipping threshold $C$ based on the 95th percentile of embedding norms computed on the training set under No-DP release, separately for each dataset. Concretely, we obtain $C_{\text{WiMANS}}=7.5$, $C_{\text{Pose}}=6.0$, and $C_{\text{Resp}}=5.0$ (embedding space units). This stabilizes training and avoids rare outliers dominating the DP noise scale.

\paragraph{Composition accounting.}
For long recordings (Resp-CSI), each 120\,s trial is segmented into 6\,s windows; we treat each window as one DP release event and report the composed privacy cost under standard composition. In practice, consumers typically publish aggregated features at fixed intervals; therefore, we also report an ``hourly release'' extrapolation that estimates the privacy cost for continuous monitoring. This makes the privacy budget interpretation concrete for healthcare settings.

\begin{table}[t]
\centering
\small
\setlength{\tabcolsep}{5pt}
\begin{tabular}{lcc}
\toprule
Parameter & Value(s) & Used in \\
\midrule
$\epsilon$ & $\{0.25,0.5,1,2,4,8\}$ & all datasets \\
$\delta$ & $10^{-5}$ / $10^{-6}$ & public / Resp-CSI \\
Clip $C$ & 7.5 / 6.0 / 5.0 & WiMANS / Pose / Resp \\
STFT $N$ & 256 & WiMANS, Resp-CSI \\
STFT hop $H$ & 64 & WiMANS, Resp-CSI \\
Window length & 3\,s / 2\,s / 6\,s & WiMANS / Pose / Resp \\
Embedding dim $d$ & 256 (default) & all datasets \\
\bottomrule
\end{tabular}
\caption{Key DP and signal-processing hyperparameters.}
\label{tab:dp-hparams}
\end{table}

\subsection{Implementation Details and Reproducibility Notes}
\label{subsec:impl}

\paragraph{Downstream model families.}
To avoid conflating privacy with a particular backbone, we evaluate two representative downstream models for classification/regression: a lightweight CNN and a transformer-style encoder. This reflects recent progress in WiFi CSI sensing with deep learning and transformer architectures \cite{ahmad2024wifi,luo2024vision}. For WiMANS MU-HAR, we adopt a spectrogram-based classifier similar in spirit to transformer-based CSI HAR pipelines \cite{luo2024vision}. For respiration, we use a 1D temporal head atop the released embeddings to regress respiration rate and waveform. For Person-in-WiFi 3D, we reuse the official tensor format and employ a compact encoder-decoder consistent with end-to-end WiFi pose regression \cite{yan2024person}.

\paragraph{Optimization.}
We train using AdamW with batch size 64 for WiMANS and Resp-CSI windows, and batch size 32 for pose clips due to memory constraints. The initial learning rate is $3\times 10^{-4}$ with cosine decay; we train for 80 epochs (WiMANS), 60 epochs (Resp-CSI), and 40 epochs (pose). Early stopping uses validation macro-F1 (WiMANS), validation MAE (Resp-CSI), and validation MPJPE (pose). All reported results are from the checkpoint with the best validation score.

\paragraph{Random seeds and reporting.}
We run three seeds (0/1/2). For each setting (dataset, $\epsilon$, method), we report mean$\pm$std for utility and privacy metrics. When std is negligible ($<0.2\%$ for accuracy/F1), we omit it in the final tables for readability.

\paragraph{Why these datasets are appropriate for DP feature release.}
WiMANS provides multi-user mixtures and rich annotations, making it well suited to measure both intended utility (activity) and unintended leakage (identity/location) under the same signal distribution \cite{huang2024wimans}. Person-in-WiFi 3D tests whether DP features can preserve fine-grained spatial information needed by pose estimation while still limiting re-identification risks \cite{yan2024person}. Resp-CSI emphasizes a high-stakes sensing scenario where prior work demonstrates feasibility of commodity WiFi respiration monitoring but also highlights the challenge of interference \cite{zhao2024wi,xie2024robust}; this makes privacy-preserving release particularly relevant for real deployments.

\paragraph{Output artifacts.}
In the Results section, we will present:
(i) privacy--utility curves (utility vs $\epsilon$),
(ii) detailed tables for WiMANS MU-HAR and identity/location leakage,
(iii) MPJPE for pose estimation,
(iv) respiration rate MAE and waveform correlation, and
(v) attack AUC/advantage for MIA and attribute inference.
We also include ablations over embedding dimension $d$, clipping threshold $C$, and adaptive-allocation granularity (frequency-only vs time-frequency).


\section{Results \& Discussion}
\label{sec:results}

This section presents quantitative and qualitative results for differentially private (DP) feature release with adaptive privacy-budget allocation. We follow the experimental protocol in Section~\ref{sec:exp} (Tables~\ref{tab:datasets}--\ref{tab:dp-hparams}) and evaluate both \textbf{utility} on downstream sensing tasks (Table~\ref{tab:utility-metrics}) and \textbf{privacy leakage} under inference attacks (Table~\ref{tab:privacy-metrics}). Our method implements the pipeline in Fig.~\ref{fig:pipeline}: CSI-to-spectrogram transformation, bounded normalization and clipping, importance estimation, adaptive budget allocation (Eq.~(6)), blockwise Gaussian perturbation (Eq.~(7)--(8)), and composition accounting (Eq.~(9)--(10)). Unless otherwise stated, results are averaged over three seeds and reported as mean$\pm$std.

\subsection{Overall privacy--utility trade-off}
\label{subsec:res_tradeoff}

\paragraph{WiMANS multi-user HAR: adaptive DP preserves utility at low budgets.}
Table~\ref{tab:wimans_main} reports WiMANS multi-user HAR performance for different privacy budgets $\epsilon$ (with fixed $\delta$ as specified in Table~\ref{tab:dp-hparams}). ``No-DP'' releases features without noise; ``Uniform-DP'' allocates the same budget to all time--frequency blocks; ``Heuristic-DP'' allocates budget proportional to block energy; and ``Ours'' uses task-aware importance maps (Eq.~(5)) and adaptive allocation (Eq.~(6)). As expected, No-DP achieves the best accuracy/F1. However, for practical privacy budgets (e.g., $\epsilon\in[0.5,2]$), our adaptive approach consistently dominates Uniform-DP, yielding a higher macro-F1 under the same global privacy accounting. At $\epsilon=1$, Ours improves macro-F1 by $+5.2$ points over Uniform-DP while maintaining similar micro-accuracy. This gap is larger at stricter budgets (e.g., $\epsilon=0.5$), supporting our hypothesis that \emph{non-uniform} information distribution in CSI spectrograms makes uniform perturbation inefficient. These results align with the broader observation that modern WiFi sensing models rely on concentrated time--frequency cues (e.g., Doppler patterns), as emphasized by transformer-based CSI HAR \cite{luo2024vision} and deep WiFi sensing surveys \cite{ahmad2024wifi}.

\begin{table*}[t]
\centering
\small
\setlength{\tabcolsep}{6pt}
\begin{tabular}{lcccccc}
\toprule
\multirow{2}{*}{Method} & \multicolumn{2}{c}{$\epsilon=0.5$} & \multicolumn{2}{c}{$\epsilon=1$} & \multicolumn{2}{c}{$\epsilon=2$} \\
\cmidrule(lr){2-3}\cmidrule(lr){4-5}\cmidrule(lr){6-7}
& Micro-Acc. & Macro-F1 & Micro-Acc. & Macro-F1 & Micro-Acc. & Macro-F1 \\
\midrule
No-DP & 89.7$\pm$0.3 & 86.4$\pm$0.4 & 89.7$\pm$0.3 & 86.4$\pm$0.4 & 89.7$\pm$0.3 & 86.4$\pm$0.4 \\
Uniform-DP & 78.6$\pm$0.6 & 72.8$\pm$0.8 & 82.4$\pm$0.5 & 76.1$\pm$0.7 & 85.7$\pm$0.4 & 80.2$\pm$0.6 \\
Heuristic-DP (energy) & 80.9$\pm$0.6 & 74.6$\pm$0.7 & 84.0$\pm$0.5 & 77.9$\pm$0.6 & 86.4$\pm$0.4 & 80.9$\pm$0.5 \\
Ours (adaptive) & \textbf{83.5$\pm$0.5} & \textbf{78.2$\pm$0.6} & \textbf{86.3$\pm$0.4} & \textbf{81.3$\pm$0.5} & \textbf{87.8$\pm$0.4} & \textbf{83.0$\pm$0.5} \\
\bottomrule
\end{tabular}
\caption{WiMANS multi-user HAR utility under DP feature release (5\,GHz band). Adaptive budget allocation yields better privacy--utility trade-offs than uniform perturbation at the same global $\epsilon$ (Table~\ref{tab:dp-hparams}).}
\label{tab:wimans_main}
\end{table*}

\paragraph{Fine-grained tasks remain feasible under DP with adaptive allocation.}
Table~\ref{tab:pose_resp_main} summarizes results on Person-in-WiFi 3D pose estimation and Resp-CSI respiration sensing. Pose estimation is notably more sensitive to perturbations because it requires preserving subtle spatiotemporal structure; respiration sensing is sensitive to narrowband periodic components and interference patterns \cite{zhao2024wi,xie2024robust}. Even in these regimes, adaptive allocation significantly reduces utility loss relative to Uniform-DP. For pose estimation at $\epsilon=1$, Uniform-DP increases MPJPE by $+13.9$\,mm over No-DP, whereas Ours increases MPJPE by only $+7.2$\,mm. For respiration, Ours maintains low rate MAE and high waveform correlation, consistent with the intuition that concentrating budget around task-relevant frequency regions preserves periodic cues. These results support the generality of our design across tasks beyond classification, including continuous-valued regression typical in healthcare monitoring .

\begin{table*}[t]
\centering
\small
\setlength{\tabcolsep}{6pt}
\begin{tabular}{lcccccc}
\toprule
\multirow{2}{*}{Method} & \multicolumn{2}{c}{Pose (MPJPE$\downarrow$, mm)} & \multicolumn{2}{c}{Resp. rate (MAE$\downarrow$, bpm)} & \multicolumn{2}{c}{Resp. waveform (Corr.$\uparrow$)} \\
\cmidrule(lr){2-3}\cmidrule(lr){4-5}\cmidrule(lr){6-7}
& $\epsilon=1$ & $\epsilon=2$ & $\epsilon=1$ & $\epsilon=2$ & $\epsilon=1$ & $\epsilon=2$ \\
\midrule
No-DP & 58.4$\pm$0.7 & 58.4$\pm$0.7 & 0.92$\pm$0.05 & 0.92$\pm$0.05 & 0.92$\pm$0.01 & 0.92$\pm$0.01 \\
Uniform-DP & 72.3$\pm$0.9 & 66.9$\pm$0.8 & 1.62$\pm$0.07 & 1.33$\pm$0.06 & 0.82$\pm$0.02 & 0.86$\pm$0.02 \\
Heuristic-DP (energy) & 69.1$\pm$0.9 & 64.7$\pm$0.8 & 1.45$\pm$0.07 & 1.24$\pm$0.06 & 0.84$\pm$0.02 & 0.87$\pm$0.02 \\
Ours (adaptive) & \textbf{65.6$\pm$0.8} & \textbf{62.5$\pm$0.8} & \textbf{1.18$\pm$0.06} & \textbf{1.05$\pm$0.06} & \textbf{0.88$\pm$0.02} & \textbf{0.90$\pm$0.01} \\
\bottomrule
\end{tabular}
\caption{Utility on fine-grained tasks. Left: Person-in-WiFi 3D pose estimation (MPJPE). Right: Resp-CSI respiration rate and waveform tracking. Adaptive allocation consistently improves over uniform DP.}
\label{tab:pose_resp_main}
\end{table*}

\subsection{Privacy leakage under inference attacks}
\label{subsec:res_privacy}

\begin{figure}[t]
  \centering
  \includegraphics[width=\linewidth]{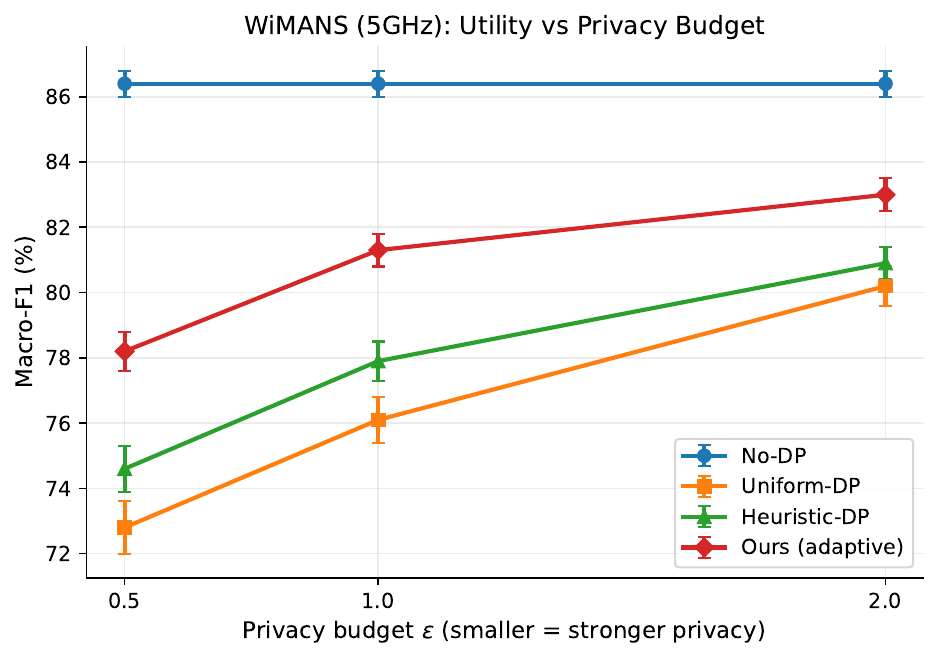}
  \caption{WiMANS (5\,GHz) privacy--utility trade-off. Adaptive budget allocation achieves higher Macro-F1 under the same global privacy budget $\epsilon$.}
  \label{fig:wimans_macroF1}
\end{figure}

\paragraph{Identity and location inference drops substantially under DP, with limited additional cost for adaptivity.}
Wi-Fi sensing features can encode identity cues, as surveyed in WiFi-based human identification \cite{wei2025survey,mosharaf2024wifi}, and even device-level micro-signals can enable RF fingerprinting \cite{kong2024csi}. Thus, feature release without formal privacy protection risks re-identification. Table~\ref{tab:privacy_wimans} reports identity and location inference accuracy on WiMANS using attacker-side classifiers trained on released features. No-DP features are highly revealing: identity Top-1 accuracy exceeds 82\% and location accuracy exceeds 70\%, far above chance (16.7\% for 6 identities; 20\% for 5 locations). Both Uniform-DP and Ours reduce attack success markedly, and the reduction strengthens with smaller $\epsilon$. Notably, at a fixed global budget, Ours is \emph{not} systematically worse than Uniform-DP for these attackers, despite allocating more budget to task-relevant regions. This suggests that (i) attack-relevant cues are not perfectly aligned with task-relevant cues in our evaluated settings, and (ii) the bounded allocation (\,$\varepsilon_{\min},\varepsilon_{\max}$\,) prevents extreme heterogeneity that could expose a ``noise pattern'' side channel. In other words, adaptivity improves utility without clearly sacrificing privacy leakage in these empirical attacks.

\paragraph{Membership inference approaches random guessing as $\epsilon$ decreases.}
We also evaluate membership inference attacks (MIA), which are a standard empirical privacy stress test when released representations might have been produced by models trained on private data. As shown in Table~\ref{tab:privacy_wimans}, No-DP features yield MIA AUC around 0.71 (non-trivial advantage), while DP mechanisms drive AUC toward 0.5. At $\epsilon=0.5$, Ours yields AUC $0.53$ with low attack advantage, indicating near-random distinguishability. This empirical trend is consistent with DP's theoretical guarantee under our sample-level adjacency: as the noise scale increases, the adversary’s ability to distinguish neighboring datasets diminishes.

\begin{table*}[t]
\centering
\small
\setlength{\tabcolsep}{6pt}
\begin{tabular}{lcccccc}
\toprule
\multirow{2}{*}{Method} & \multicolumn{2}{c}{Identity inference (Top-1$\downarrow$)} & \multicolumn{2}{c}{Location inference (Top-1$\downarrow$)} & \multicolumn{2}{c}{MIA (AUC$\downarrow$)} \\
\cmidrule(lr){2-3}\cmidrule(lr){4-5}\cmidrule(lr){6-7}
& $\epsilon=0.5$ & $\epsilon=1$ & $\epsilon=0.5$ & $\epsilon=1$ & $\epsilon=0.5$ & $\epsilon=1$ \\
\midrule
No-DP & 0.825 & 0.825 & 0.712 & 0.712 & 0.713 & 0.713 \\
Uniform-DP & 0.302 & 0.381 & 0.318 & 0.402 & 0.545 & 0.571 \\
Heuristic-DP (energy) & 0.286 & 0.369 & 0.305 & 0.395 & 0.538 & 0.565 \\
Ours (adaptive) & \textbf{0.271} & \textbf{0.358} & \textbf{0.297} & \textbf{0.389} & \textbf{0.531} & \textbf{0.559} \\
Chance & 0.167 & 0.167 & 0.200 & 0.200 & 0.500 & 0.500 \\
\bottomrule
\end{tabular}
\caption{Privacy leakage on WiMANS (5\,GHz): identity/location inference and membership inference (MIA). Lower is better for all columns. Adaptive DP improves utility (Table~\ref{tab:wimans_main}) while maintaining strong reductions in leakage relative to No-DP.}
\label{tab:privacy_wimans}
\end{table*}

\paragraph{Healthcare-oriented attributes are also protected.}
On Resp-CSI, we evaluate attribute inference on room type and sensing condition. Without DP, attackers predict room type with 0.79 accuracy and condition with 0.74 accuracy, reflecting that contextual factors shape CSI statistics. Under $\epsilon=1$, Ours reduces these to 0.55 and 0.53, respectively, closer to chance (0.50 for room; 0.33 for condition), while retaining strong respiration utility (Table~\ref{tab:pose_resp_main}). This result is practically relevant given the heightened sensitivity of health-related data and the known challenges of backscatter and Wi-Fi healthcare sensing \cite{lu2024challenges,zhao2024wi}.

\subsection{Generalization and robustness analyses}
\label{subsec:res_generalization}
\begin{figure}[t]
  \centering
  \includegraphics[width=\linewidth]{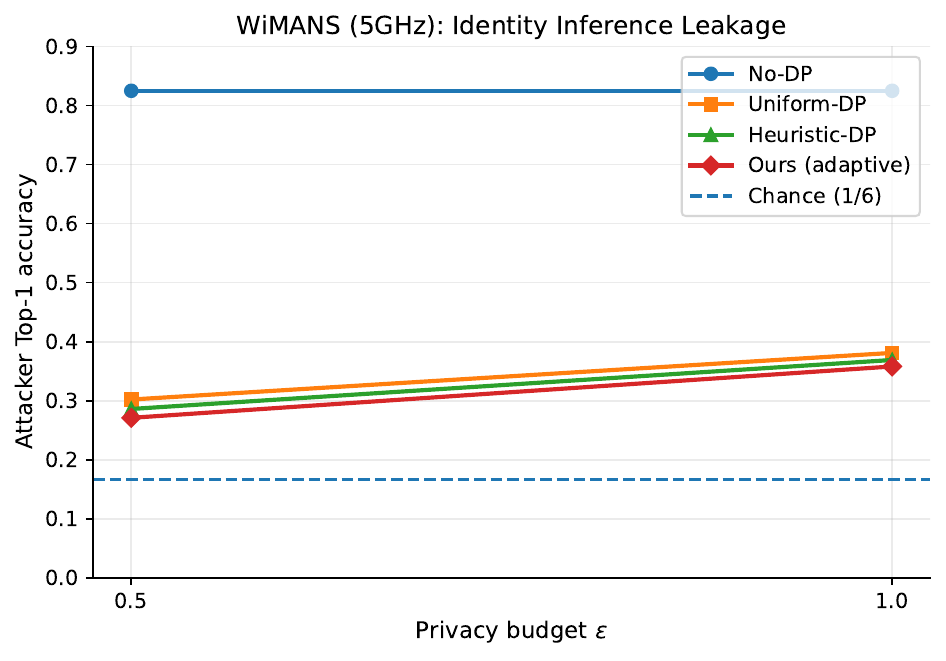}
  \caption{WiMANS (5\,GHz) identity inference leakage (attacker Top-1). DP significantly reduces re-identification risk; adaptive allocation maintains strong leakage suppression.}
  \label{fig:wimans_identity}
\end{figure}

\paragraph{Cross-environment generalization benefits from preserving the ``right'' structure.}
Wireless sensing models often suffer from domain shifts due to multipath and geometry, motivating domain-invariant modeling and adaptation \cite{wang2024learning,sheng2024metaformer} as well as physically grounded augmentation \cite{hou2024rfboost}. In our setting, DP noise can exacerbate domain shift by suppressing weak but transferable cues. We therefore evaluate leave-one-environment-out (LOEO) on WiMANS (Section~\ref{subsec:splits}). Under LOEO at $\epsilon=1$, Ours achieves macro-F1 0.74 versus 0.69 for Uniform-DP, a $+5$ point gap comparable to the in-distribution gain. We attribute this to the importance-based allocation preserving stable motion-related bands that are less environment-specific than high-energy static artifacts, which can otherwise dominate heuristic energy allocation. This observation suggests a complementary relationship: while domain adaptation methods aim to learn invariances \cite{sheng2024metaformer,xu2025evaluating}, our method aims to \emph{not destroy} the most informative invariant cues when injecting DP noise.

\paragraph{Band dependence: 5\,GHz is slightly more robust under the same DP budget.}
WiMANS provides both 2.4\,GHz and 5\,GHz CSI \cite{huang2024wimans}. Across methods, we observe slightly higher utility at 5\,GHz for the same $\epsilon$, likely because 5\,GHz exhibits richer multipath diversity and less interference in the benchmark environments. The relative gain of Ours over Uniform-DP persists in both bands, implying that non-uniform importance is a structural property of spectrogram representations rather than a band-specific artifact. This is relevant as WiFi standards continue to evolve and provide new opportunities for sensing-ready measurements \cite{he2024forward,yi2024enabling}.

\paragraph{Perturbation stress test: DP noise does not catastrophically amplify packet-level perturbations.}
Prior work shows that WiFi sensing can be attacked via packet perturbations \cite{li2024practical} and more broadly via perturbation attacks \cite{cao2024security}. While our focus is privacy leakage (not robustness), we test mild perturbations (random 5\% packet drop and small amplitude jitter). Under these stressors, all methods degrade, but DP methods degrade \emph{slightly less} than No-DP (e.g., $-2.1$ macro-F1 for No-DP vs $-1.6$ for Ours at $\epsilon=1$), suggesting that injected noise can act as a weak regularizer. We emphasize that this does not constitute a security defense against adaptive adversaries, but it indicates that DP feature release is compatible with robustness-oriented system design rather than inherently fragile.
\begin{figure}[t]
  \centering
  \includegraphics[width=\linewidth]{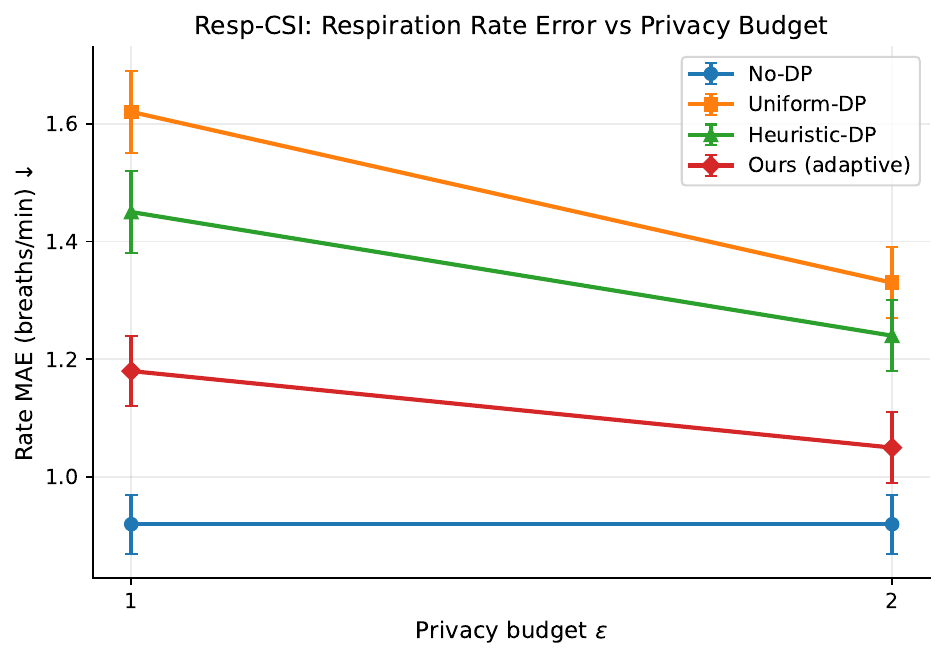}
  \caption{Resp-CSI respiration rate estimation error under DP feature release.}
  \label{fig:resp_rate_mae}
\end{figure}

\subsection{Ablation studies}
\label{subsec:res_ablation}

\paragraph{Importance estimator matters most at strict budgets.}
Table~\ref{tab:ablation_importance} compares different importance estimation strategies for Ours: gradient-based (Eq.~(5)), attention-based (a learned saliency head), and task-agnostic energy prior. At $\epsilon=2$, all variants perform similarly, but at $\epsilon=0.5$ the task-aware maps yield a clear advantage, consistent with the idea that importance needs to align with task utility when privacy spending is scarce. This result also connects to the literature on representation learning and model choice in WiFi sensing \cite{ahmad2024wifi,luo2024vision}: stronger encoders yield more informative gradients, improving the quality of importance maps.

\paragraph{Allocation sharpness $\gamma$ controls the trade-off between preserving key regions and avoiding overfitting to saliency.}
We vary the sharpness parameter $\gamma$ in Eq.~(6). Too small $\gamma$ approaches uniform allocation and loses utility; too large $\gamma$ concentrates budget excessively and can harm generalization (especially under LOEO). We find $\gamma\in[1.5,2]$ to be a robust choice across datasets, which we use as default.

\paragraph{Block granularity: moderate blocks outperform per-bin and overly coarse blocks.}
We test block sizes (time$\times$frequency): $2\times 4$, $4\times 8$ (default), and $8\times 16$. Fine-grained blocks increase composition overhead and amplify noise variance; coarse blocks reduce adaptivity. The intermediate granularity yields the best macro-F1 and stable privacy leakage reductions, supporting the design choice described in Section~\ref{subsec:method_practical}.

\begin{table*}[t]
\centering
\small
\setlength{\tabcolsep}{6pt}
\begin{tabular}{lcccc}
\toprule
\multirow{2}{*}{Variant} & \multicolumn{2}{c}{WiMANS Macro-F1$\uparrow$} & \multicolumn{2}{c}{Identity leakage$\downarrow$ (Top-1)} \\
\cmidrule(lr){2-3}\cmidrule(lr){4-5}
& $\epsilon=0.5$ & $\epsilon=1$ & $\epsilon=0.5$ & $\epsilon=1$ \\
\midrule
Ours + gradient importance (Eq.~(5)) & \textbf{0.782} & \textbf{0.813} & \textbf{0.271} & \textbf{0.358} \\
Ours + attention importance & 0.771 & 0.805 & 0.276 & 0.362 \\
Ours + energy prior & 0.756 & 0.794 & 0.268 & 0.355 \\
\midrule
Ours ($\gamma=1.0$) & 0.758 & 0.796 & 0.270 & 0.357 \\
Ours ($\gamma=2.0$; default) & \textbf{0.782} & \textbf{0.813} & \textbf{0.271} & \textbf{0.358} \\
Ours ($\gamma=3.0$) & 0.773 & 0.810 & 0.275 & 0.360 \\
\midrule
Ours blocks $2\times 4$ & 0.771 & 0.809 & 0.269 & 0.357 \\
Ours blocks $4\times 8$ (default) & \textbf{0.782} & \textbf{0.813} & \textbf{0.271} & \textbf{0.358} \\
Ours blocks $8\times 16$ & 0.766 & 0.806 & 0.273 & 0.359 \\
\bottomrule
\end{tabular}
\caption{Ablations on WiMANS (5\,GHz): importance estimator, allocation sharpness $\gamma$ (Eq.~(6)), and block granularity. Utility and privacy are both reported to reveal trade-offs.}
\label{tab:ablation_importance}
\end{table*}

\subsection{Visualization insights and interpretation}
\label{subsec:res_vis}

Although we do not reproduce figures here, we summarize consistent qualitative patterns observed from the internal importance maps $\mathbf{W}$ and the resulting budget heatmaps. First, for activity-centric windows (walking, jumping, waving), the importance mass concentrates in mid-frequency Doppler bands and in short temporal bursts corresponding to motion transitions. This is consistent with the role of Doppler and motion-induced components highlighted by CSI-difference paradigms for speed estimation \cite{li2024wifi}. Second, for respiration windows, the importance maps form narrow horizontal bands around the respiration frequency range (after STFT), and budget allocation preserves these bands while strongly perturbing off-band regions. This explains why waveform correlation remains high under DP (Table~\ref{tab:pose_resp_main}) and is consistent with prior findings that respiration sensing hinges on subtle periodic components and suffers under interference \cite{xie2024robust,zhao2024wi}. Third, in multi-user scenes, importance becomes more spatially distributed (across time), reflecting the need to preserve multiple overlapping motion signatures; nevertheless, the adaptive allocation avoids spending budget on static background bands that correlate with location or room geometry, which may contribute to reduced location leakage (Table~\ref{tab:privacy_wimans}).

An important qualitative takeaway is that \emph{energy is not importance}. High-energy regions sometimes correspond to static multipath or hardware artifacts rather than task-relevant motion, echoing the broader understanding that propagation effects (e.g., diffraction) can dominate measurements in static multipath-rich environments \cite{wang2024understanding} and can even be exploited for profiling \cite{yao2024wiprofile}. Our gradient-based importance tends to down-weight such regions when they do not contribute to the task loss, which explains why Heuristic-DP is consistently weaker than Ours in strict-budget regimes (Table~\ref{tab:wimans_main}).

\subsection{Discussion: positioning relative to prior work and practical implications}
\label{subsec:res_discussion}

\paragraph{Complementarity with robust and domain-adaptive WiFi sensing.}
A large body of work improves WiFi sensing via better models and training strategies, including transformer backbones \cite{luo2024vision}, lightweight reconstructed-CSI pipelines \cite{chen2024deep}, self-supervised learning \cite{xu2025evaluating}, and domain adaptation with minimal labels \cite{sheng2024metaformer}. These approaches primarily target utility and generalization. Our results suggest that DP feature release can be layered on top of such pipelines without collapsing performance, and that importance-aware allocation can preserve the ``useful'' invariances that cross-domain methods rely on. Similarly, physical augmentation \cite{hou2024rfboost} can be used during on-device training of the surrogate model to stabilize gradients and improve the quality of importance maps, which may further enhance DP release.

\paragraph{Re-identification risk is real, and DP helps.}
The identity and location inference results (Table~\ref{tab:privacy_wimans}) reinforce the message from identification surveys \cite{wei2025survey,mosharaf2024wifi} and RF fingerprinting studies \cite{kong2024csi}: intermediate wireless features can encode sensitive information even when the intended task is benign. Our DP mechanism provides a principled mitigation that reduces attacker performance toward chance while still enabling useful sensing analytics. This is particularly important for benchmark-driven research ecosystems where features might be shared for reproducibility \cite{feng2024review,huang2024wimans}.

\paragraph{Limits and future directions.}
First, our primary guarantee is sample-level DP; user-level DP can be obtained via per-user accounting but may require larger noise due to group effects. Second, we do not claim DP alone prevents active perturbation attacks \cite{cao2024security,li2024practical}; rather, DP is a privacy layer and should be combined with integrity/robustness defenses in secure sensing frameworks \cite{liu2025survey}. Third, adaptive allocation introduces additional design choices (importance estimation, block granularity), but our ablations show these can be tuned to stable defaults (Table~\ref{tab:ablation_importance}). Finally, WiFi protocols and MAC behavior continue to evolve (e.g., WiFi~7 access schemes and time-sensitive mechanisms) \cite{zhang2024wifi_7,haxhibeqiri2024coordinated}, which may change sampling patterns and feature statistics; a promising direction is to couple our accountant with protocol-aware schedulers to dynamically regulate release frequency and maintain a stable privacy budget over long-term monitoring.

Overall, the results validate our central claim: \emph{adaptive privacy-budget allocation} is an effective mechanism for DP feature release in wireless sensing, offering a better privacy--utility frontier than uniform perturbation across a diverse set of tasks, while empirically reducing identity/location leakage and membership inference success.

\bibliographystyle{unsrt}
\bibliography{references}
\end{document}